\documentclass[journal]{IEEEtran}

\normalsize
\usepackage{mathtools}
\usepackage{float}
\usepackage{dblfloatfix}

\usepackage{cite}
\usepackage{amsmath,amssymb,amsfonts}
\usepackage{algorithmic}
\usepackage{graphicx}
\usepackage{textcomp}
\usepackage{mathtools}
\usepackage{xcolor}
\usepackage{xspace}
\usepackage{acronym}
\usepackage{multirow}
\usepackage{tabularx}
\usepackage{todonotes}
\usepackage[ruled, lined, commentsnumbered, linesnumbered]{algorithm2e}
\usepackage{caption}
\usepackage{soul}
\usepackage{subcaption}
\usepackage{amsmath}
		\usepackage{pgfplots}
		\usepackage{color}
		\usepackage{tikz}
		\usepackage{multicol}
		\usepackage{hyperref}
\usepackage{amssymb} 
\usepackage{pifont} 

\newcommand{\cmark}{\ding{51}}
\newcommand{\xmark}{\ding{55}}

 \pgfplotsset{compat=1.17}
 
 \usepackage[letterpaper, left=0.65in, right=0.65in, bottom=1.03in, top=0.8in]{geometry}
\usepackage{changes}
\usepackage{lipsum}
\definechangesauthor[name={Abdullatif}, color=red]{revised/added}
\definechangesauthor[name={Abdullatif}, color=red]{Revised/Added}
\definechangesauthor[name={Abdullatif2}, color=red]{deleted}
\definechangesauthor[name={Abdullatif3}, color=red]{revised}
\definechangesauthor[name={Abdullatif4}, color=red]{moved} 
\definechangesauthor[name={Abdullatif4}, color=red]{abbreviated}
  
\begin{document}
\title{The Role of Deep Learning in Advancing Proactive Cybersecurity Measures for Smart Grid Networks: A Survey}
\author{Nima Abdi, Abdullatif Albaseer,~\IEEEmembership{Member,~IEEE,}
        Mohamed Abdallah,~\IEEEmembership{Senior Member,~IEEE,}
\thanks{Nima Abdi, Abdullatif Albaseer, and Mohamed Abdallah are with the Division of Information and Computing Technology, College of Science and Engineering, Hamad Bin Khalifa University, Doha, Qatar 
\{niab52126,aalbaseer, moabdallah\}@hbku.edu.qa}
}

\maketitle
\begin{abstract}
As smart grids {(SG)} increasingly rely on advanced technologies like sensors and communication systems for efficient energy generation, distribution, and consumption, they become enticing targets for sophisticated cyber-attacks. These evolving threats demand robust security measures to maintain the stability and resilience of modern energy systems. While extensive research has been conducted, a comprehensive exploration of proactive cyber defense strategies utilizing Deep Learning (DL) in {SG} remains scarce in the literature. This survey bridges this gap, studying the latest DL techniques for proactive cyber defense.
The survey begins with an overview of related works and our distinct contributions, followed by an examination of SG infrastructure. Next, we classify various cyber defense techniques into reactive and proactive categories. A significant focus is placed on DL-enabled proactive defenses, where we provide a comprehensive taxonomy of DL approaches, highlighting their roles and relevance in the proactive security of {SG}. Subsequently, we analyze the most significant DL-based methods currently in use. Further, we explore Moving Target Defense, a proactive defense strategy, and its interactions with DL methodologies. We then provide an overview of benchmark datasets used in this domain to substantiate the discourse.{ This is followed by a critical discussion on their practical implications and broader impact on cybersecurity in Smart Grids.}
The survey finally lists the challenges associated with deploying DL-based security systems within {SG}, followed by an outlook on future developments in this key field. 
\end{abstract}
 

\section{Introduction}\label{sec_intro}
A smart grid (SG) is an advanced electricity delivery system that leverages digital technology to enhance the power grid's efficiency, reliability, and sustainability. The SG facilitates the integration of renewable energy sources such as solar and wind power, electric vehicles, and energy storage systems, promoting eco-friendly electricity generation. It harnesses the potential of cutting-edge technologies such as sensors, communication networks, and advanced control systems. This utilization enables real-time monitoring and control over the electric power grid \cite{intro_28}. The role of SG is crucial in adapting to the dynamic conditions presented by distributed generation and renewable energy, with its advanced technologies facilitating the bidirectional flow of electrical energy. Its objective is to establish a flexible system capable of accommodating power demand, supply, and storage options associated with the distribution system.
The seamless exchange of data and communication among consumers, power utilities, and system operators in {SG} is made possible by implementing smart meters (SMs) in residences, buildings, and industrial facilities, a key component of the Advanced Metering Infrastructure (AMI). SMs observe consumers' electricity usage and transmit detailed readings, either through wireless \cite{intro_6} or wired \cite{intro_7} communications methods. The data is then conveyed to the utility company for power quality monitoring and pricing management \cite{intro_8}. This can help consumers monitor and manage their energy consumption offering the potential for cost savings and a reduced carbon footprint` \cite{intro_3}. Supervisory Control and Data Acquisition (SCADA) system monitors and manages the transmission network. However, despite the benefits the SG offers, it is important to address the security and privacy issues associated with its implementation  \cite{intro_9}.

Integrating digital technology into the electric power grid introduces new vulnerabilities that cybercriminals can exploit, which result in disruption of power supply, theft of sensitive data, and damage to critical infrastructure \cite{Attacks_A_DoS_1, Attacks_A_Buffer_1, Attacks_C_1}. Cyberattacks against {SG} can take many forms, including customer data theft, the failure of cyber systems, compromising communication instruments, spreading false data and malware, and eavesdropping \cite{AE4, LSTM1}.

Ensuring the security of the SG is vital to protecting the electric power system's reliability and public safety, as cybersecurity breaches can disrupt the power supply and damage critical infrastructure. Implementing robust security measures, including incident response and recovery plans, is essential. Within the SG, communication protocols such as IEC 60870-5-104 and DNP3 are widely utilized, providing standardized data transfer and control commands \cite{intro_16, intro_17}. However, these protocols lack critical security measures like authentication and integrity protection, leaving them vulnerable to attacks such as Man-in-the-middle (MITM) \cite{intro_18, MITM_1}, denial-of-service (DoS) \cite{intro_19}, and spoofing \cite{intro_20}. Such vulnerabilities also extend to other common SG protocols like DNP3 \cite{intro_21, intro_22, CS_non_AI_1}. The extended lifetimes of power grid field devices, like protection devices and PLCs, further exacerbate the potential for these attacks, underlining the need for comprehensive security measures.

Some energy companies may face practical constraints when upgrading to newer models that support protocols like IEC 61850, enabling communication with intelligent electrical devices (IEDs) \cite{intro_26}. The implementation of IEC 61850 introduces vulnerabilities that necessitate attention, potentially expanding the attack surface \cite{intro_27}. Modern security mechanisms must seamlessly integrate with grid operation devices without disrupting the power supply, but some may lack the computational capacity for additional security functions \cite{intro_24}. Obsolete protocols pose a security challenge, even during device upgrades \cite{intro_25}. SG cybersecurity measures include secure network architectures, communication protocols, access controls, and intrusion detection systems (IDS) to enhance security \cite{intro_3}.

Cyberdefense organizations employ reactive and proactive strategies. Reactive cyber security involves responding to and mitigating attacks with countermeasures such as IDS, incident response, and patch management \cite{intro_3}. Proactive cyber defense emphasizes threat intelligence analysis, security awareness training, and continuous monitoring to prevent cyber threats before they begin. {To further enhance cybersecurity, machine learning (ML) and deep learning (DL) techniques are employed. ML refines the performance of countermeasures by training on large datasets \cite{CS_ML_1}. DL, a more advanced subset of ML, uses artificial neural networks for real-time detection and response to threats. It automates feature extraction in artificial intelligence (AI), reducing the need for manual intervention and resource allocation compared to traditional ML methods \cite{17}. This approach enables DL to independently tackle complex problems, effectively manage unstructured data like images and audio, and suit big data contexts with its ability to handle large inputs and diverse outputs \cite{22}. Additionally, DL surpasses traditional methods in cybersecurity with greater precision, lower false negative rates, and higher classification accuracy \cite{23}. Its proficiency in managing complex, non-linear data significantly boosts its performance in network anomaly detection.  DL not only predicts and prevents future threats by analyzing historical data to identify patterns and trends, \cite{AE3,DRL5} but also adapts to the dynamic nature of cybersecurity, making it an outstanding choice for advanced, data-intensive applications \cite{24}. It is worth mentioning that our paper specifically emphasizes DL-based techniques for SG security, recognizing their superior capability in handling high-dimensional data and complex patterns. This focus aligns with the established consensus in the literature (as in~\cite{14, 17}) on the advanced efficacy of DL over traditional ML methods.}

\begin{figure}[t]
\centering
\includegraphics[width=\linewidth]{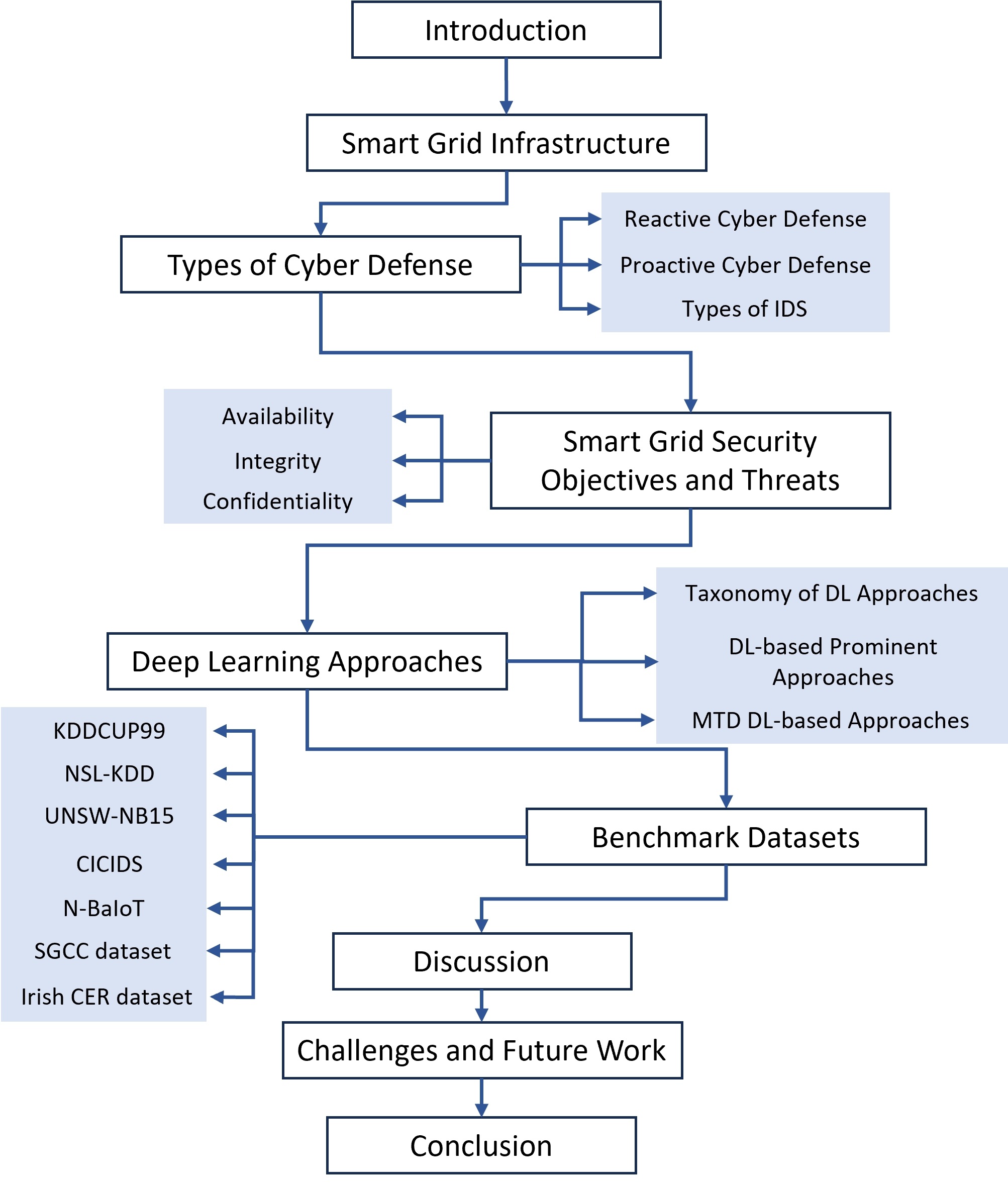}
\caption{{Paper organization structure.}}
\label{fig:paper_organization}
\end{figure}

\subsection{{Related Work} and Contribution}
Cyberattacks on {SG} encompass various attack methods, exploiting vulnerabilities with different intentions and strategies. Ding et al. \cite{intro_10} studied attack strategies and their characteristics and explored the potential of using AI and blockchain to enhance detection and prevention in {SG}. Other studies, such as those cited in \cite{intro_11, intro_12, intro_13}, focus on ML and DL applications in cybersecurity, particularly in SG and IDS. \cite{intro_11} offers a comprehensive analysis of the challenges and prospects of ML in SG cybersecurity. \cite{intro_12} compares ML methods, including DL, in IDS, emphasizing DL's benefits in handling complex data. \cite{intro_13} aims to enhance detection accuracy, reduce false alarms, and detect unknown attacks through a taxonomy framework that categorizes IDS literature based on ML and DL approaches. Ferrag et al. \cite{intro_14} discussed seven distinct DL models' effectiveness, assessing them using specific datasets and evaluation metrics. \cite{intro_15} expands on DL applications in SG, including Federated Learning (FL), edge intelligence, and distributed computing, highlighting significant applications like energy forecasting and fault detection. Several authors have analyzed MTD techniques, including classification and categorization, examining methodologies, algorithms, and evaluation methods, but with a noticeable lack of emphasis on DL techniques \cite{intro_16_MTD, intro_17_MTD, intro_18_MTD}.

Previous surveys in DL-based cybersecurity and MTD have significantly contributed, but a research gap exists in proactive security in SG environments. While many techniques, classification methods, and evaluations have been studied, little attention has been given to DL-based proactive security measures specific to {SG}. Therefore, further research is needed to better understand DL-based proactive security solutions for SG systems. This will give the researcher an overview of the existing techniques and where to use them.  To differentiate between our survey (i.e., contributions) and the existing ones, Table \ref{tab:comparison} presents a detailed comparison. Our study aims to explore current DL approaches in SG proactive security, providing insights into the strengths, weaknesses, and comparisons of primary DL approaches. We also investigate emerging challenges and future advancements in SG and DL models.
 
\begin{table*}[t]
\centering
\caption{Comparison of Contributions}
\begin{tabularx}{\textwidth}{|X|c|c|c|c|c|c|c|c|}
\hline
\textbf{Ref} & \multicolumn{1}{p{1.3cm}|}{\centering \textbf{Attack Strategies}} & \multicolumn{1}{p{1.3cm}|}{\centering \textbf{AI/Blockchain}} & \multicolumn{1}{p{1.8cm}|}{\centering \textbf{ML/DL Applications}} & \multicolumn{1}{p{1.8cm}|}{\centering \textbf{SG-specific Challenges}} & \multicolumn{1}{p{1.8cm}|}{\centering \textbf{Benchmarking Datasets}} & \multicolumn{1}{p{1.3cm}|}{\centering \textbf{DL Models}} & \multicolumn{1}{p{1.8cm}|}{\centering \textbf{MTD Techniques}} & \multicolumn{1}{p{1.8cm}|}{\centering \textbf{DL-based Proactive Sec.}} \\ \hline
\cite{intro_10} & \cmark & \cmark & \xmark & \xmark & \xmark & \xmark & \xmark & \xmark \\ \hline
\cite{intro_11} & \xmark & \xmark & \cmark & \cmark & \xmark & \xmark & \xmark & \xmark \\ \hline
\cite{intro_12} & \xmark & \xmark & \cmark & \xmark & \cmark & \xmark & \xmark & \xmark \\ \hline
\cite{intro_13} & \xmark & \xmark & \cmark & \xmark & \xmark & \xmark & \xmark & \xmark \\ \hline
\cite{intro_14} & \xmark & \xmark & \cmark & \xmark & \cmark & \cmark & \xmark & \xmark \\ \hline
\cite{intro_15} & \xmark & \xmark & \cmark & \xmark & \xmark & \cmark & \xmark & \xmark \\ \hline
\cite{intro_16_MTD} & \xmark & \xmark & \xmark & \xmark & \xmark & \xmark & \cmark & \xmark \\ \hline
\cite{intro_17_MTD} & \xmark & \xmark & \xmark & \xmark & \xmark & \xmark & \cmark & \xmark \\ \hline
\cite{intro_18_MTD} & \xmark & \xmark & \xmark & \xmark & \xmark & \xmark & \cmark & \xmark \\ \hline
Our Survey & \xmark & \xmark & \cmark & \cmark & \cmark & \cmark & \cmark & \cmark \\ \hline
\end{tabularx}
\label{tab:comparison}
\end{table*}

As for the remainder of the paper, section \ref{sec: Overview SG} provides a general overview of SG infrastructure. Section \ref{CD} provides types of cyber defense techniques, such as reactive and proactive approaches.
Section \ref{sec: security objectives and attacks} describes the security objectives and potential threats in {SG}.
Section \ref{sec: DL_approaches}  will provide a general overview and prominent DL models and applications utilizing DL and MTD in SG.
Section \ref{sec: Benchmark Datasets} examines benchmark datasets widely utilized in {SG}. 
{Section \ref{sec: Discussion} presents a comprehensive discussion on DL and MTD's application in SG, providing a deeper understanding of their impact and effectiveness.}
Section \ref{sec: Challenges} explores challenges in implementing DL for SG cybersecurity and suggests research directions.
Lastly, the study is concluded in section \ref{sec: Conclusion}. Fig. \ref{fig:paper_organization} shows the organization of our paper.

\section{SG Infrastructure}
\label{sec: Overview SG}
A SG represents a smart evolution of the conventional physical grid aiming for secure, efficient, and eco-friendly transmission of power demand by utilizing advanced sensing, communication, and decision-making technology. This section introduces the SG infrastructure’s key components and functionalities.

\subsubsection{SCADA system}
SCADA systems enhance industrial automation by enabling optimized and efficient operations. Comprising a control server, communication channels, and field devices, they monitor, control, and transmit signals to electrical devices. The control center processes, stores, and displays data on an HMI. However, increased interconnectivity and remote access introduce security vulnerabilities, making SCADA systems susceptible to real-time communication breaches \cite{SG_Inf_2}.

\subsubsection{Distribution automation}
\begin{figure*}[!t]
\centering
\includegraphics[width=0.7\linewidth]{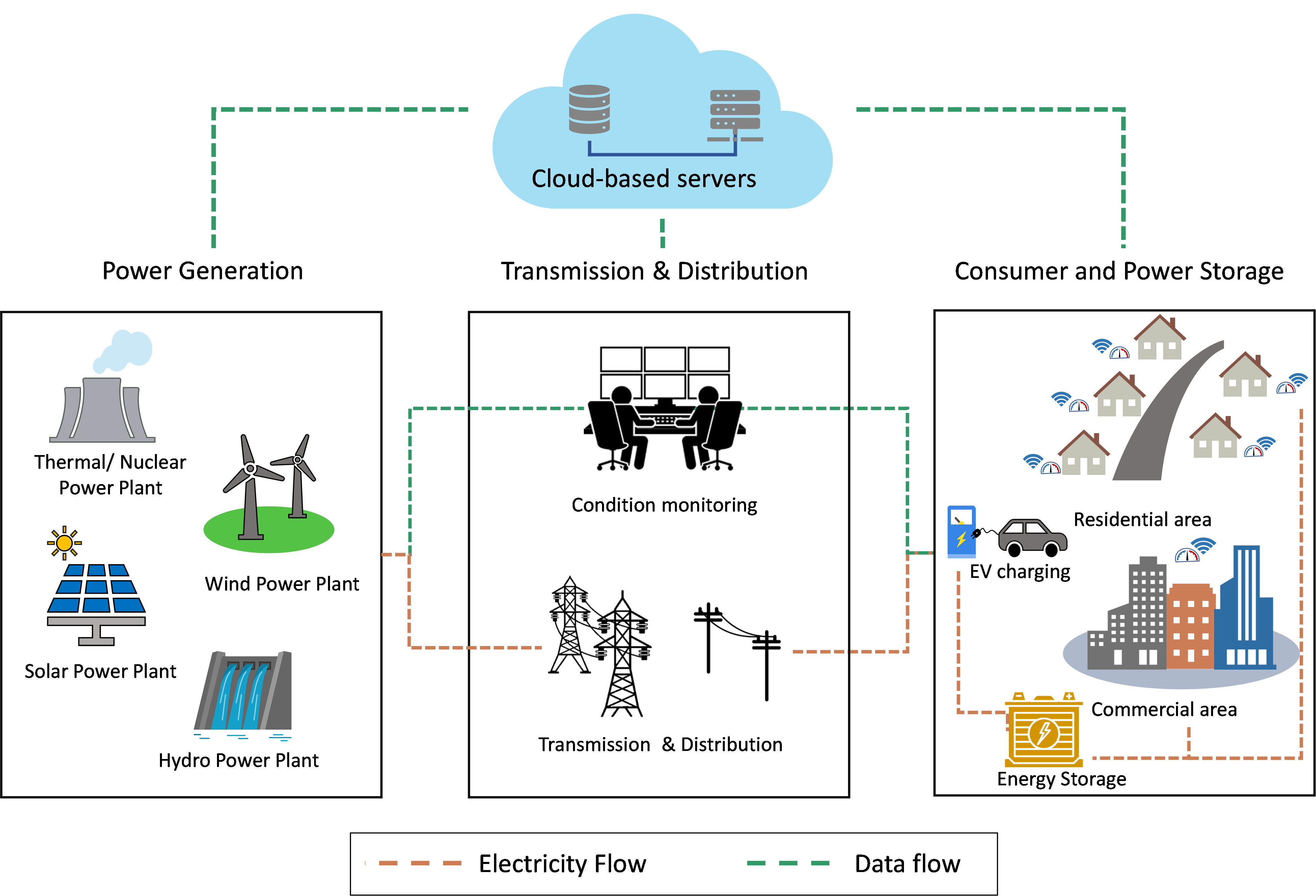}
\caption{SG Architecture}
\label{fig:SG_architecture}
\end{figure*}

Distribution automation monitors network conditions, detecting faults and reducing outage duration. Fig. \ref{fig:SG_architecture} illustrates the integration within SG architecture, showcasing energy generation, distribution, and consumption.

\subsubsection{Distributed energy resources (DERs)}
DERs are compact power sources crucial in meeting electricity needs. These solutions, including storage systems and renewable technologies, support power demand. Distributed generation encompasses technologies like wind turbines, solar panels, and batteries, but the integration also requires managing vast volumes of data \cite{SG_Inf_3, SG_Inf_9}.

\subsubsection{Energy storage systems (ESSs)}
ESSs address the imbalance between variable energy generation and fluctuating demand \cite{SG_Inf_4}. Users can reduce energy costs by storing and discharging excess energy to meet increased load. ESSs effectively alleviate the disparity between production and consumption, ensuring renewable energy sources' reliability \cite{SG_Inf_5}.

\subsubsection{Advanced Metering Infrastructure}
AMI represents a comprehensive system enhancing {SMs} by enabling bidirectional communication, accurate billing, and energy consumption management. It goes beyond simple meter reading, incorporating devices and systems for advanced data management, communication, and analysis \cite{SG_Inf_6}.
\subsubsection{Microgrids}
Microgrids are networks of connected loads and dispersed energy sources, controlling the grid using renewable energy sources. They provide better power quality, enhancing the grid efficiency and reducing environmental impact \cite{SG_Inf_12}.

\subsubsection{Internet of Things {(IoT)}}
The SG can integrate devices, sensors, and systems via IoT, resulting in a network of interconnected components. It acquires and transmits real-time data from various sources. However, integrating IoT introduces potential vulnerabilities and security risks, such as susceptibility to cyberattacks \cite{SG_Inf_10, SG_Inf_11}.

\subsubsection{Edge computing}
The SG establishes an efficient energy transmission network, incorporating various IoT devices. Edge computing deploys servers near data sources, enabling localized services and optimized data analysis and processing tasks from IoT devices, improving overall efficiency within the power grid \cite{SG_Inf_7}.

\section{Cyber {Defense in SG}}
\label{CD}
Cyber attacks pose a significant challenge for organizations, including {SG}. Reactive cybersecurity measures, while foundational, sometimes require support from a proactive defense approach. This section examines both reactive and proactive cyber defense approaches in SG.

\subsection{Reactive Cyber {Defense}}
Reactive cybersecurity focuses on detecting and countering threats, often using {IDS} that alert organizations after incidents have occurred.

\subsubsection{Rule-Based Approaches}
Initially, rule-based systems were prevalent, operating by established attack patterns. However, their limitations led to more advanced techniques such as knowledge-based and anomaly-based methods in {SG} \cite{CD_R_3, CD_R_4}.

\subsubsection{Knowledge-based (signature-based) IDS}
This approach stores predefined intrusion patterns, achieving high detection accuracy but failing to identify novel attacks. It requires frequent updates in dynamic settings \cite{CD_R_5, CD_R_6, CD_R_7}.

\subsubsection{Anomaly-based techniques}
Anomaly-based IDS identifies malicious activities by analyzing deviations from normal behavior. Effective in identifying unseen attacks, it can lead to a high rate of false alarms, and defining the threshold of normal behavior is challenging \cite{CD_R_8, CD_R_9, CD_R_10, CD_R_13}.

\subsubsection{Deception}
Deception techniques, such as honeypots, have gained popularity. {These techniques are categorized based on their level of interaction with attackers. They serve as a reactive strategy for identifying and addressing ongoing attacks and also enable the collection of forensic evidence \cite{CD_R_11, CD_R_12, CD_A_1, CD_A_2, CD_A_3}.}

\begin{figure}[t]
\centering
\includegraphics[width=0.7\linewidth]{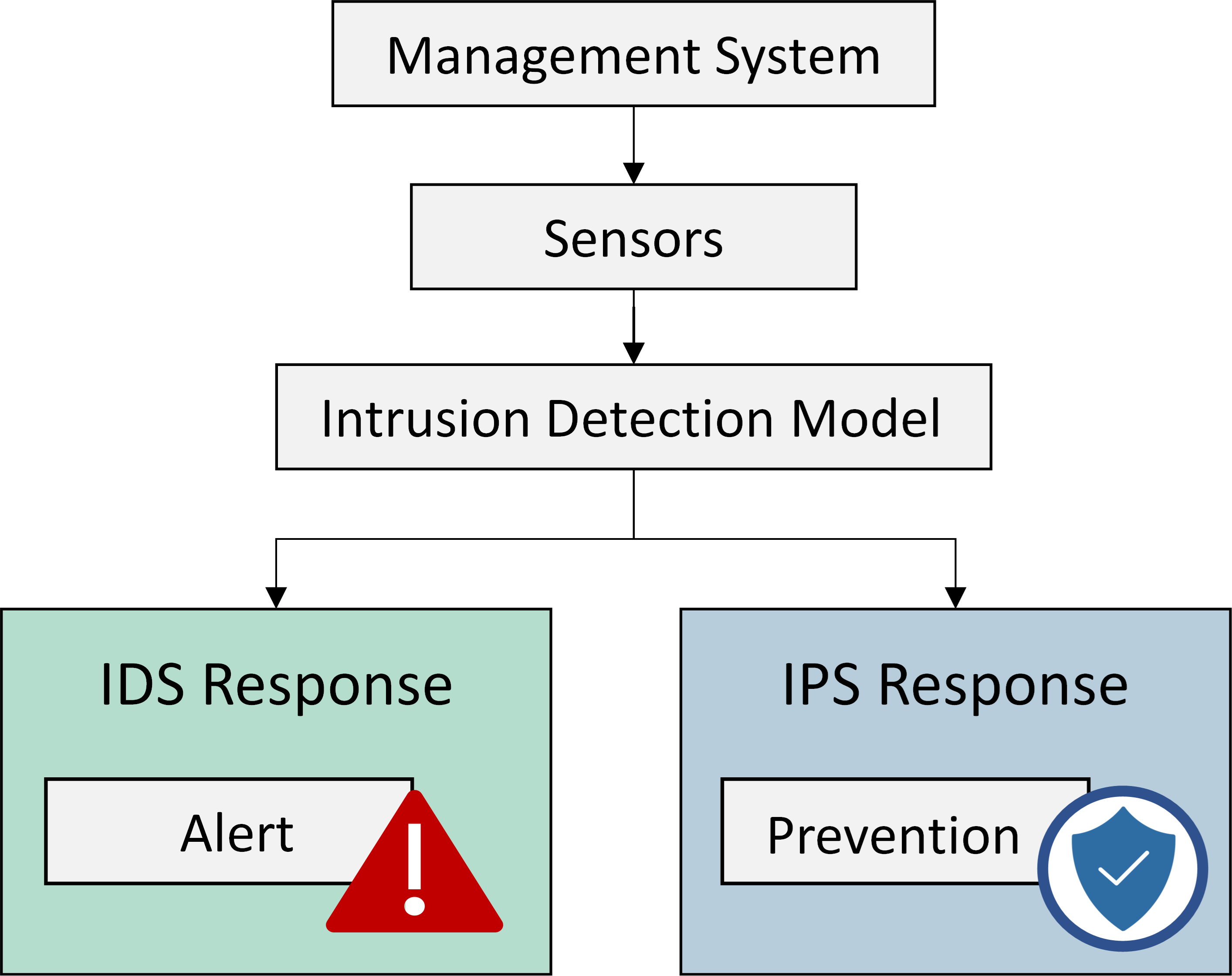}
\caption{IDS and IPS Framework}
\label{fig:IDS_IPS}
\end{figure}
\subsection{Proactive Cyber {Defense}}
Proactive security measures enhance resilience and defend against online attacks, ensuring the integrity and reliability of SG operations. These measures enable effective threat identification, prompt accident response, and anticipating attackers' activities. Let's examine proactive cyber defense strategies that improve SG security. 

\subsubsection{Intrusion prevention systems}
Intrusion prevention systems (IPS) involve conducting intrusion detection and making efforts to halt potential incidents that have been identified. IPS often incorporate IDS functionality as part of their framework and may utilize a combination of these techniques to detect and mitigate potential malicious behavior within a network \cite{CD_A_4, CD_A_5}. While IPS and IDS are commonly discussed together, it is important to note that they are distinct tools with different primary focuses. See Fig. \ref{fig:IDS_IPS} for an illustration depicting the relationship between the IDS and IPS response.

\subsubsection{Encryption}
Encryption secures data by changing its structure, preventing unauthorized access and false information injection. The data becomes incomprehensible without the encryption key, thereby discouraging unauthorized exploitation. This method safeguards against eavesdropping or tampering, ensuring confidentiality and integrity \cite{CD_R_1}. Garg et al. introduced an authentication scheme for {SMs}, utilizing various cryptographic techniques \cite{CD_R_1}. Syed et al. \cite{CD_R_2} proposed a privacy-preserving DL model for {SG} applications, using homomorphic encryption. While crucial, encryption must be complemented by other defensive measures for comprehensive cyber threat protection.

\subsubsection{Moving Target Defense (MTD)}
MTD is a proactive security approach that constantly modifies the system's attack surface, making it difficult for attackers to exploit vulnerabilities \cite{intro_16_MTD, intro_17_MTD}. This includes changing IP addresses \cite{CD_P_1}, diversifying operating systems \cite{CD_P_2}, implementing redundancy measures \cite{CD_P_3}, and employing deceptive techniques like honeypots \cite{CD_P_4}. By doing so, MTD diminishes the effectiveness of pre-attack reconnaissance, hindering the attackers' intelligence gathering \cite{intro_16_MTD, intro_17_MTD}. Unlike other cyber deception techniques, MTD focuses on system modification rather than misinformation \cite{CD_P_5}. It has been integrated with decision-making approaches like game theory \cite{CD_P_6, CD_P_7, CD_P_8}, Markov Decision Process \cite{CD_P_9, CD_P_10, CD_P_11, CD_P_12}, and {DL} schemes \cite{CD_P_13, CD_P_14, CD_P_15, CD_P_16, CD_P_17, CD_P_18}.

\begin{figure}[t]
\centering
\includegraphics[width= 0.7\linewidth]{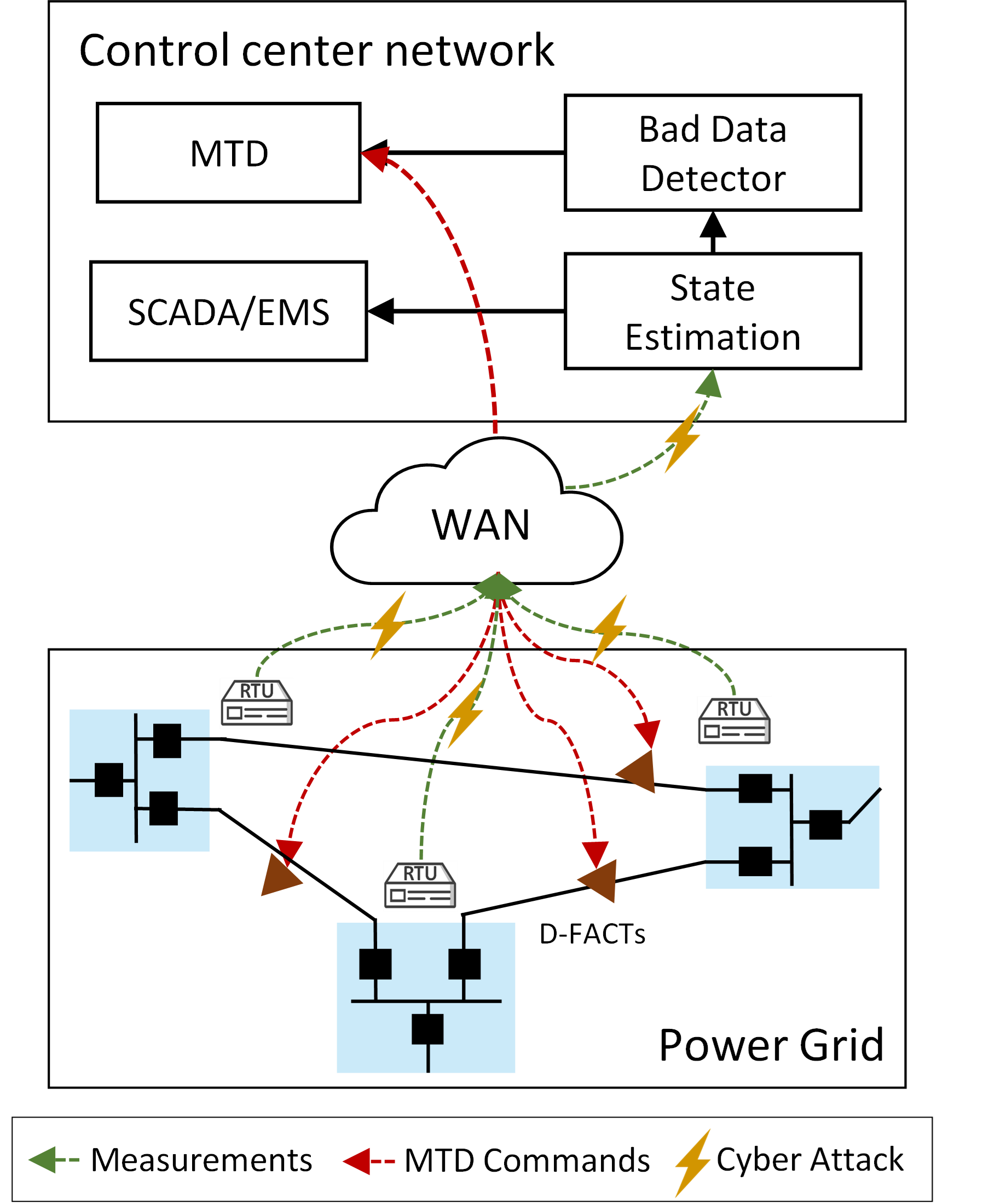}
\caption{Applying MTD to Mitigate Injection Attacks in Smart Grids \cite{intro_4}.}
\label{fig:MTD}
\end{figure}

MTD was recently explored for enhancing the defense of state estimation (SE) in cyber-physical power systems \cite{intro_4, CD_P_19}. Implementing MTD increases the complexity for attackers by introducing uncertainty and elevating the attack cost. This is achieved through the random modification of measurements and topology concerning line admittances, and the utilization of Distributed Flexible AC Transmission System (D-FACTS) devices to perturb line parameters \cite{CD_P_21, CD_P_20}, as shown in Fig. \ref{fig:MTD}.
This randomization must be carefully managed to maintain system stability and reliability \cite{intro_4, CD_P_19}. The full placement strategy using D-FACTS devices provides a more robust defense but at a higher cost \cite{intro_16_MTD}. A comparison of SG cyber defense approaches is provided in Table \ref{tab:smart_grid_cyber_defence}.

\begin{table*}[]
\centering
\caption{Comparison of Cyber {Defense} Approaches for {SG}}
\begin{tabular}{|p{2cm}|p{5.5cm}|p{2.5cm}|p{2.5cm}|p{3.5cm}|}
\hline
\textbf{Cyber {Defense} Approach} & \textbf{Description} & \textbf{Advantages} & \textbf{Limitations} & \textbf{SG Application} \\ \hline
Reactive Cyber {Defense} & This approach focuses on autonomously detecting threats and identifying potential vulnerabilities, which allows for rapid notifications to organizations about cybersecurity incidents, facilitating prompt, efficient response and assisting post-incident measures. & Provides continuous protection, reducing the need for constant human intervention. Efficient in minimizing exposure to known threats. & Lacks proactive threat hunting capability, hence can be ineffective against novel or evolving threats. It only responds once a threat has occurred. & A foundational part of any SG cybersecurity strategy. Effective for dealing with known threats but can struggle with unidentified or novel threats. \\ \hline
Proactive Cyber {Defense} & Proactive {Defense} in SG setting is about being one step ahead of potential threats. This includes measures such as threat hunting, which actively searches for and counters threats before they can impact the SG. & Enables the anticipation and prevention of threats before they affect the SG. Enables early detection through proactive activities. & This approach can be resource-intensive and requires specialized expertise. Requires deep knowledge of the constantly evolving threat landscape. & Despite being resource-intensive, it's the most comprehensive defense method. It’s crucial for ensuring the highest level of security in {SG} by pre-empting threats before they occur. \\ \hline
\end{tabular}
\label{tab:smart_grid_cyber_defence}
\end{table*}
 
\subsubsection{DL-based proactive measures}
DL-based anomaly detection enables early detection and prevention of cyber threats. These approaches allow organizations to take timely actions and enhance their overall cyber security by analyzing large volumes of data. There are various methods by which DL can improve physical and cybersecurity in {SG}:
    \paragraph{Network security} leveraging DL algorithms makes it possible to identify and block malicious traffic on the network, which can prevent cyber attacks like DoS attacks that disrupt services \cite{DL_AE2}. 
         \paragraph{Access control} DL algorithms can detect and prevent unauthorized access to SG assets, enhancing access control and protecting against cyber attacks compromising the electric power system's security \cite{Others_LSTM_CNN5, CNN1}.
          \paragraph{Intrusion detection} DL algorithms can identify and respond to cyber threats in real-time, mitigating the risks of cyber attacks that aim to compromise the security of the electric power system \cite{AE3, AE5}.
         \paragraph{Alert prioritization} DL algorithms can analyze and prioritize security alerts, enabling security teams to concentrate their resources on the most critical threats.
         \paragraph{Physical security} DL-based techniques can analyze video footage from security cameras to detect unusual behavior or physical tampering with power grid assets, which can aid in protecting against physical threats to the electric power system \cite{CD_P_22}.
\subsection{Types of IDS}
     \paragraph{Host-based intrusion detection systems (HIDS)} HIDS monitor and analyze internal system activity and behavior at the single host level. They detect internal attacks like malware infections and unauthorized access attempts, collecting data from host event logs. IDS alerts system administrators, end users, or both when anomalous behavior is detected, allowing restrictions on user requests \cite{IDS_types1}.
    \paragraph{Network-based Intrusion Detection System (NIDS)} NIDS monitors network traffic to protect against network-borne threats. It analyzes data using network components like switches, routers, or sensors, identifying intrusions targeting nodes or devices, including port scanning, DoS, and {Distributed DoS (DDoS)} attacks. NIDS logs information in system records \cite{IDS_types2}. 
    
    \paragraph{Hybrid IDS} It integrates network-based and host-based systems to enhance intrusive activity detection in distributed environments. This improves security for hosts and networks, but managing both detection functions within a single IDS can be challenging \cite{IDS_types1}. 

\section{SG Security Objectives and Threats}
\label{sec: security objectives and attacks}
\begin{figure*}[!t]
\centering
\includegraphics[width=0.7\linewidth]{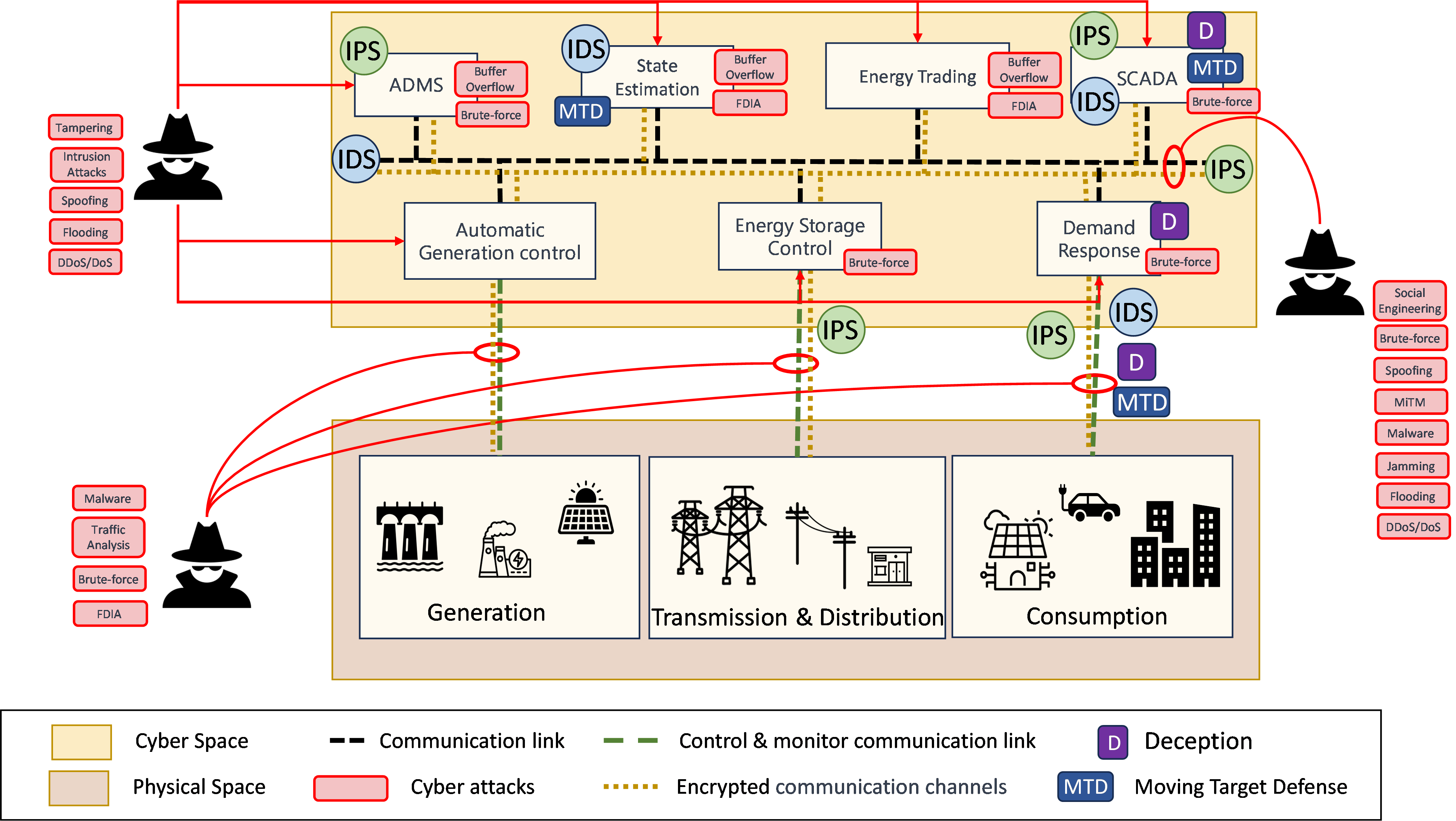}
\caption{Illustration of cyber-physical attacks defense mechanisms in SG.}
\label{fig:SG_attack_defense}
\end{figure*}

The deployment of {SG} requires integrating numerous devices, including {SM}, sensors, and renewable energy sources that communicate by sharing control commands and information. However, this increased interaction and integration makes the SG more susceptible to potential attacks, threatening its stability and safety. Therefore, the US National Institute of Standards and Technology (NIST) created a framework to thwart cyberattacks to reduce the occurrence and impact of cyber-attacks on crucial SG infrastructure. The SG must thus integrate security mechanisms, continuous monitoring, and proactive risk mitigation strategies to achieve the following NIST security objectives. Fig. \ref{fig:SG_attack_defense} visually illustrates potential attack locations within the  SG system, presenting a comprehensive overview of attack and defense aspects. It showcases the areas where various cyber-physical attacks can be simulated and analyzed, including the defense mechanisms surveyed in Section \ref{CD}.

\begin{figure}[t]
\centering
\includegraphics[width=0.8\linewidth]{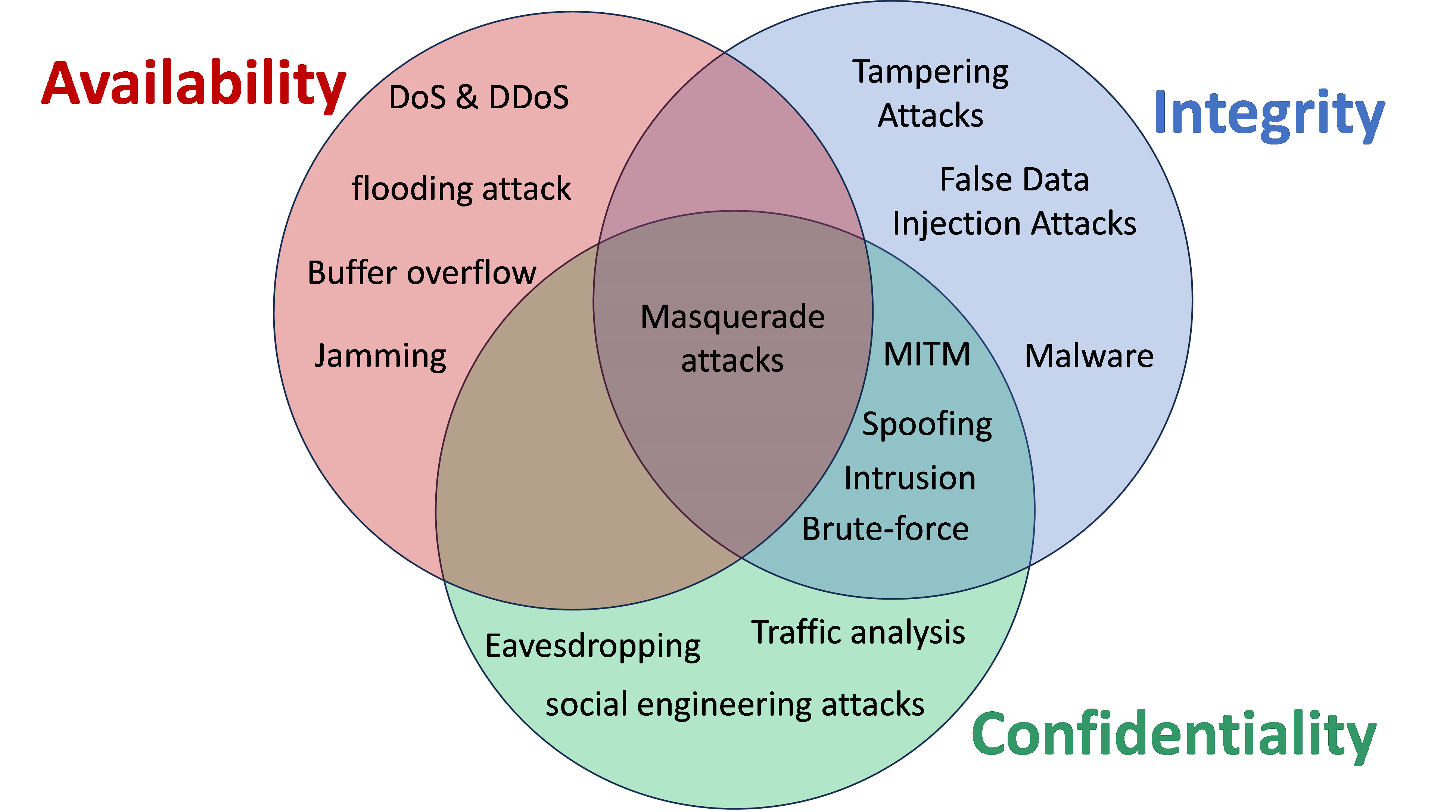}
\caption{Illustration of attacks intersection on security objectives: Integrity, Confidentiality, and Availability.}
\label{fig:attacks}
\end{figure}

The objectives of security include availability, integrity, and confidentiality. {Masquerade attacks, in particular, pose a significant threat to all these objectives, unlike various other attacks that may compromise only one or two.} The attacker can bypass security measures and gain access to sensitive information by exploiting valid credentials or unauthorized knowledge of the target's identity, undermining the system's overall security. 
Fig. \ref{fig:attacks} depicts various attacks targeting the security objectives, compromising the system's overall security.

\subsection{Availability} 
SG security focuses on availability, ensuring reliable, timely customer information access. This contributes to the power supply's efficiency by providing customers' uninterrupted power supply.
    \paragraph{DoS and DDoS attacks} DoS and DDoS attacks disrupt wireless communication networks in {SG} by overwhelming systems with excessive traffic or requests, preventing legitimate users from accessing them. An example of such attacks is flooding attacks, which can cause system crashes or render the network unresponsive \cite{Attacks_A_DoS_1}.
    \paragraph{Jamming attacks} These attacks disrupt communication systems by continuously broadcasting signals or using deceptive techniques, compromising availability and hindering communication among authorized devices \cite{Attacks_A_Jamming_2, Attacks_A_Jamming_3}. Attackers employ various types of jammers, including constant, random, deceptive, and reactive jammers, to interfere with the communication channel and block the transmission and reception of information \cite{Attacks_A_Jamming_1}.
    \paragraph{Buffer overflow attacks} The attacks here occur when a vulnerability in a system or application within the grid allows an attacker to flood a buffer with excessive data. This overflow can lead to critical data corruption, disrupt system operations, and create the potential for cascading failures, ultimately resulting in system crashes \cite{Attacks_A_Buffer_1}.
  
\subsection{Integrity} 
Cybercriminals may attempt to gain unauthorized access to SG assets, such as control systems, to compromise the security of the electric power system. Integrity refers to protecting against unauthorized change or destruction of information or resources to maintain the accuracy and efficiency of data flow throughout the system operation. Examples of integrity attacks include:
 \paragraph{Malware} Cybercriminals may use malware, such as viruses and ransomware, to compromise the security of SG systems and disrupt power supply \cite{datasets_15}.
     \paragraph{MITM attacks} MITM attacks allow attackers to interrupt and alter the communication between devices, such as {SM} and the energy management system. Modifying the transmitted information might interfere with electricity distribution and harm the grid infrastructure \cite{Attacks_A_Jamming_3}.
    \paragraph{{False Data Injection (FDI)} attacks} FDI attacks aims to manipulate the data in a network or system; this includes altering sensor readings of critical infrastructure systems like power grids. By doing so, the system believes it is functioning properly when it is not, causing significant harm or damage \cite{DRL3}.
    \paragraph{Spoofing} Spoofing involves impersonating a legitimate party to deceive others into disclosing sensitive information through packet exchange with network devices. Attackers can introduce false or misleading data, potentially causing incorrect decisions and transmitting unauthorized content \cite{Attacks_I_2}.
    \paragraph{An intrusion attack} The attacks herein occur when an adversary exploits network vulnerabilities to gain unauthorized access to the network's nodes. This unauthorized access allows attackers to tamper with files, inject malicious code, or manipulate configurations, potentially compromising the reliability and usability of the system \cite{Attacks_I_1}.
    \paragraph{Brute-force attacks} These attacks compromise the integrity of the { SG control system} by targeting unauthorized access. In SG, the central control system and various devices rely on secure protocols with encryption for communication. However, the brute-force attack undermines the system's security measures, exploiting vulnerabilities and jeopardizing the integrity of the control system \cite{Attacks_I_3}.
    \paragraph{Tampering attacks} The attacks here involve unauthorized modification of data or systems \cite{Attacks_C_1}.
 \subsection{Confidentiality} 
 Confidentiality attempts to secure personal privacy by ensuring that only legitimate individuals can access sensitive information while such information is not exposed to others without proper authorization.
    \paragraph{Eavesdropping} Eavesdropping allows unauthorized attackers to access confidential information exchanged between authorized parties by monitoring communication channels without engaging in packet exchange or active participation. Through eavesdropping, attackers can gain sensitive data, exploit vulnerabilities, and access confidential information \cite{Attacks_C_1}.
    \paragraph{MITM attack:} The attacker positions themselves between two communicating parties, enabling them to intercept the data flow. This interception allows the attacker to monitor the conversation and illicitly obtain confidential data covertly \cite{Attacks_A_Jamming_3}.
    \paragraph{An intrusion attack} An intrusion attack compromises SG data confidentiality by unauthorized access to communication channels and central control systems. The attacker gains access to sensitive information, causing privacy violations and vulnerability exposure, potentially exploiting in future attacks \cite{Attacks_I_1}.
    \paragraph{Brute-force attacks} Brute-force attacks on SG can compromise its confidentiality by systematically attempting various combinations of credentials to gain unauthorized access to sensitive data.
    \paragraph{Spoofing attacks} Spoofing attacks compromise information security, privacy, and trust, allowing unauthorized access.
    \paragraph{Social engineering} attackers use persuasion and communication techniques to gain victims' trust, obtaining private information like passwords and personal identification numbers for unauthorized access to target systems. Common methods include pretexting, phishing, and tailgating attacks \cite{Attacks_C_2}.
    \paragraph{Traffic analysis attacks} Traffic analysis attacks monitor message timing and size to determine hosts' identities and locations. The attacker extracts communication patterns between nodes by sniffing and examining messages, aiming to manipulate the SG network \cite{Attacks_C_3}.

\begin{figure*}[!t]
  \begin{subfigure}{0.3\textwidth}
    \includegraphics[width=\linewidth]{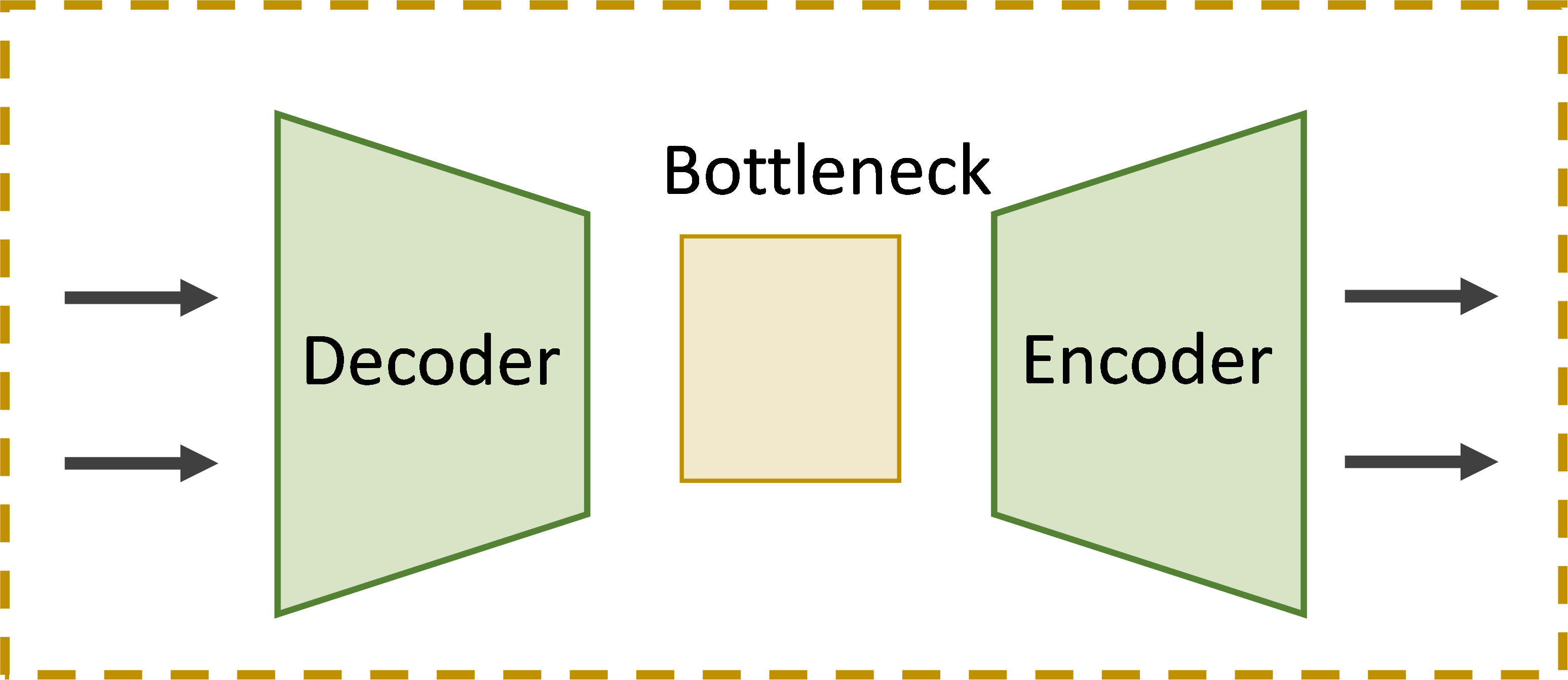}
    \caption{Autoencoder}
    \label{fig:subfig1}
  \end{subfigure}
  \hfill
  \begin{subfigure}{0.3\textwidth}
    \includegraphics[width=\linewidth]{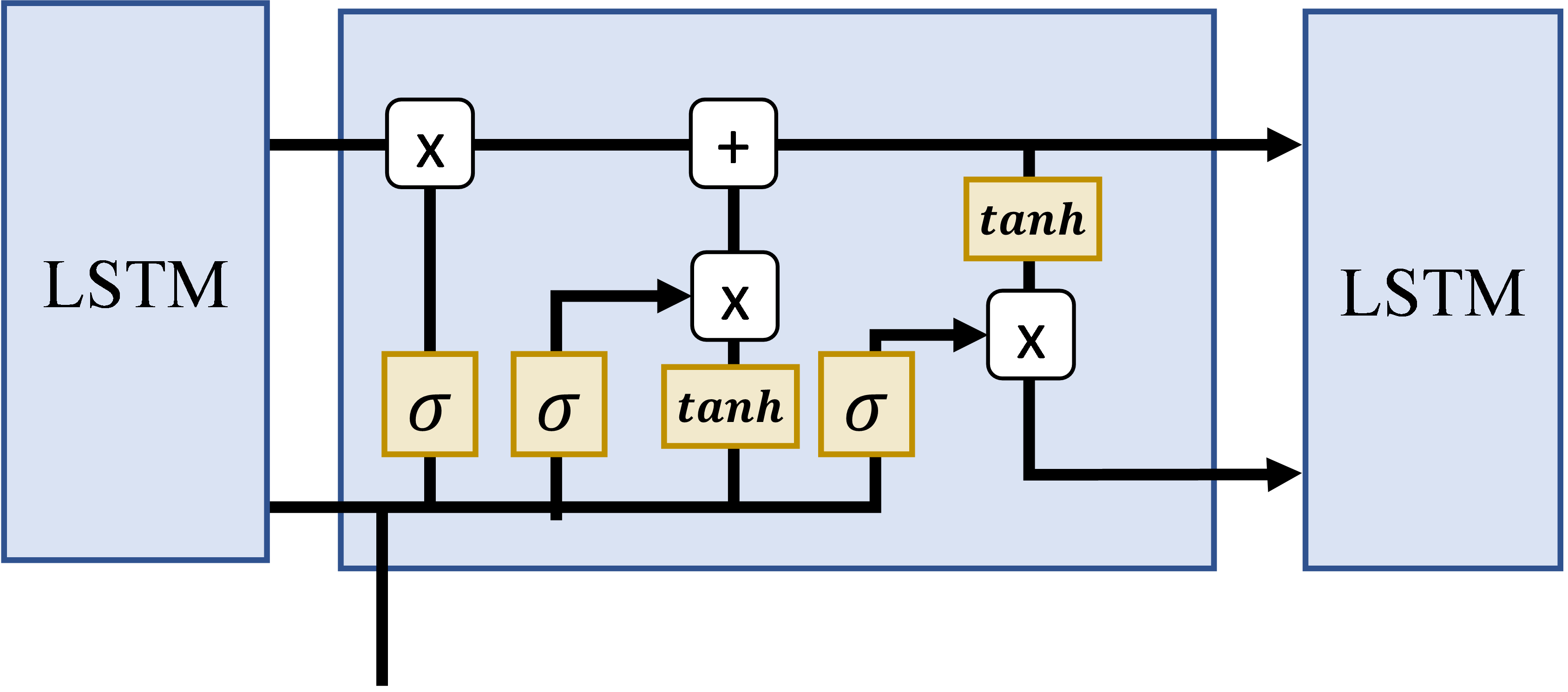}
    \caption{LSTM}
    \label{fig:subfig2}
  \end{subfigure}
  \hfill
  \begin{subfigure}{0.3\textwidth}
    \includegraphics[width=\linewidth]{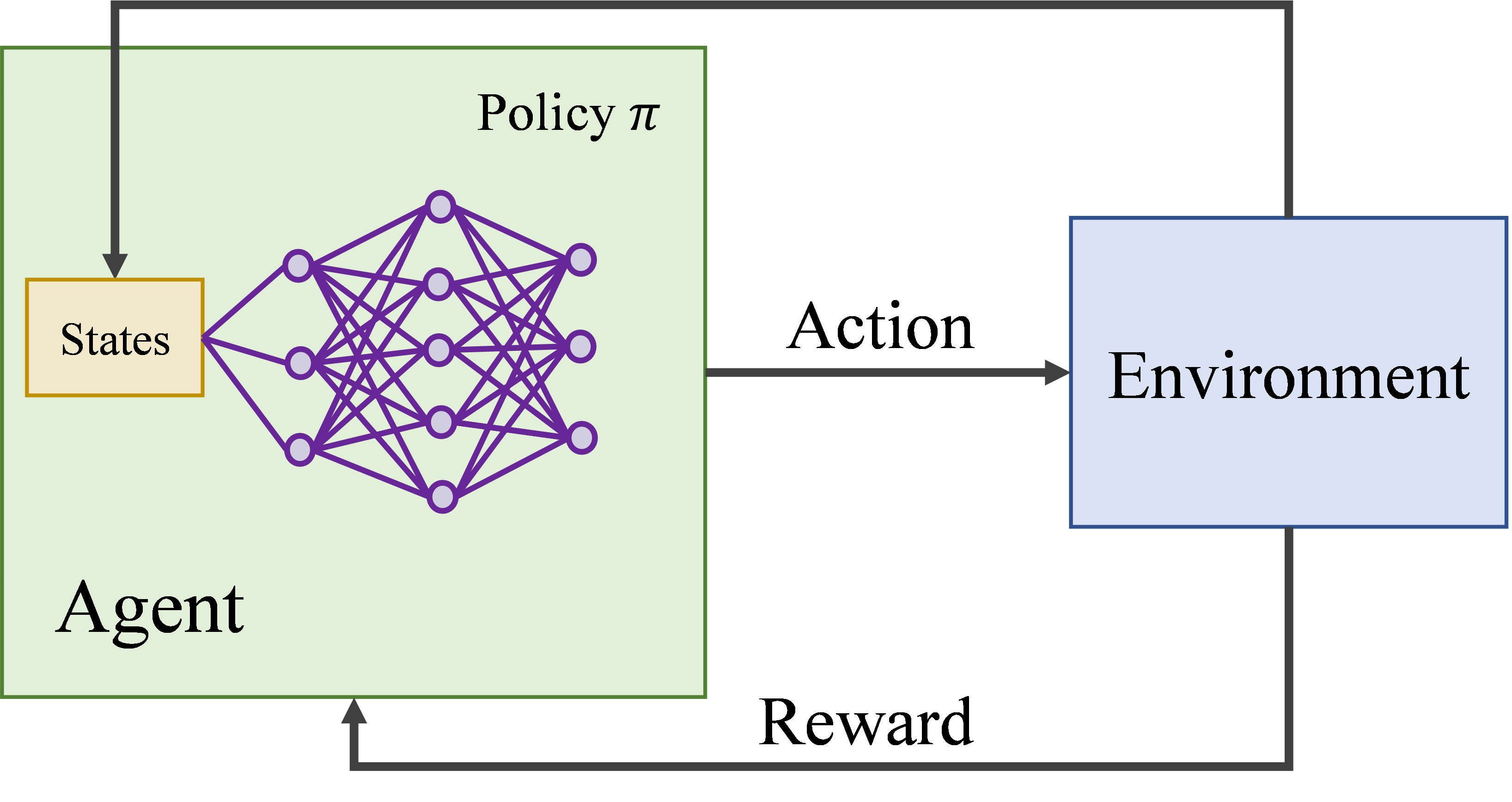}
    \caption{DRL}
    \label{fig:subfig3}
  \end{subfigure}

  \vspace{0.5cm}

  \begin{subfigure}{0.3\textwidth}
    \includegraphics[width=\linewidth]{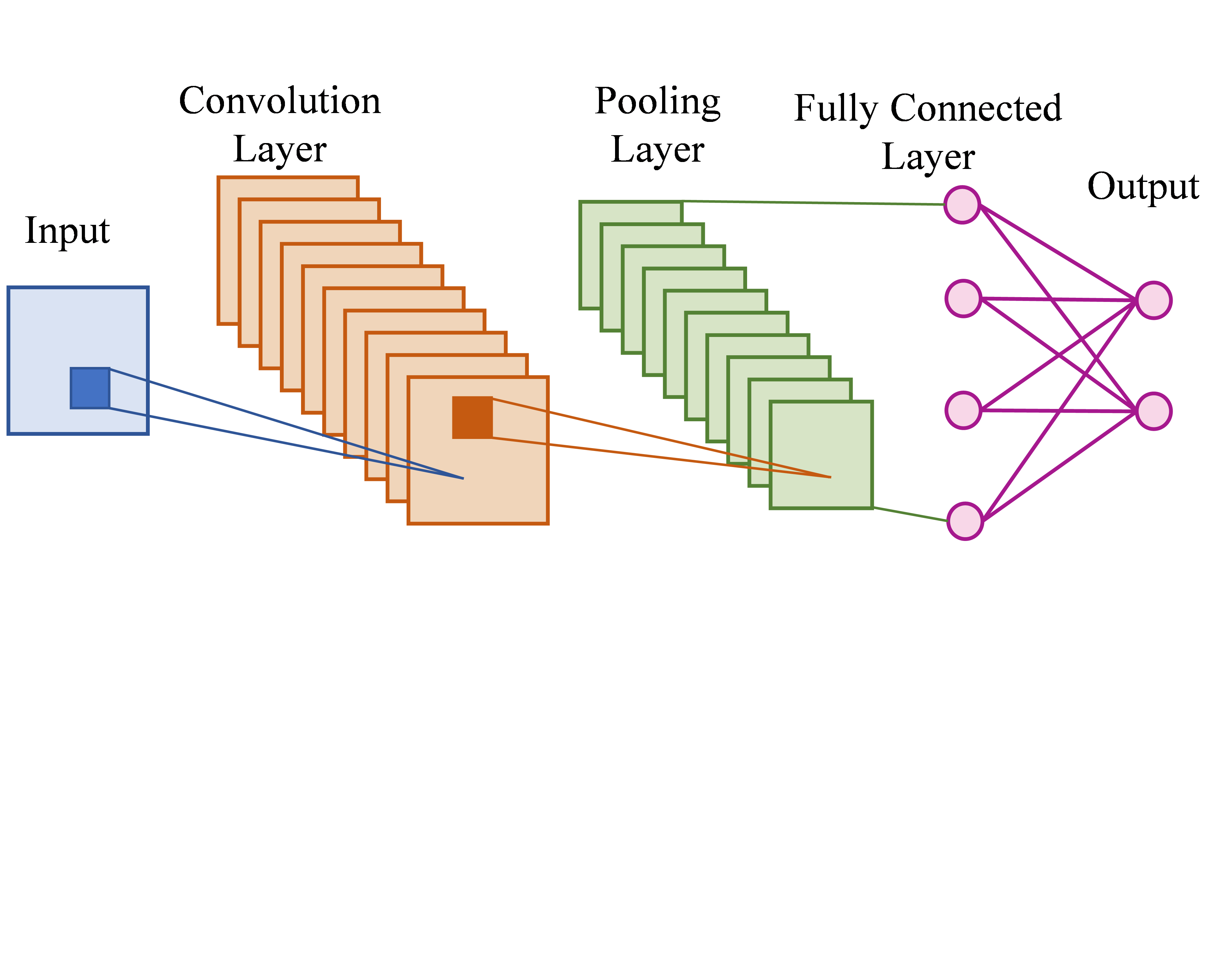}
    \caption{CNN}
    \label{fig:subfig4}
  \end{subfigure}
  \hfill
  \begin{subfigure}{0.3\textwidth}
    \includegraphics[width=\linewidth]{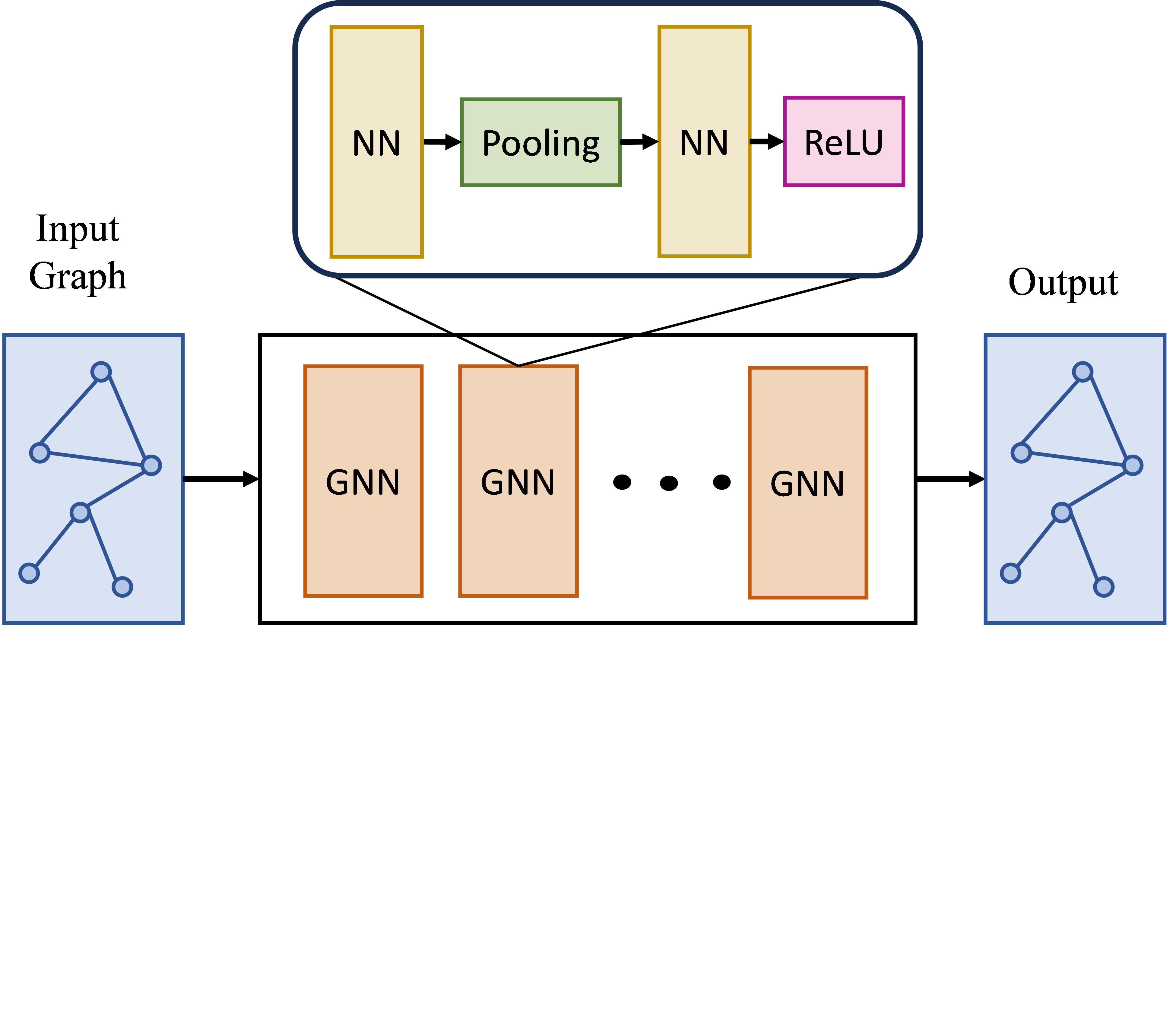}
    \caption{GNN}
    \label{fig:subfig5}
  \end{subfigure}
  \hfill
  \begin{subfigure}{0.3\textwidth}
    \includegraphics[width=\linewidth]{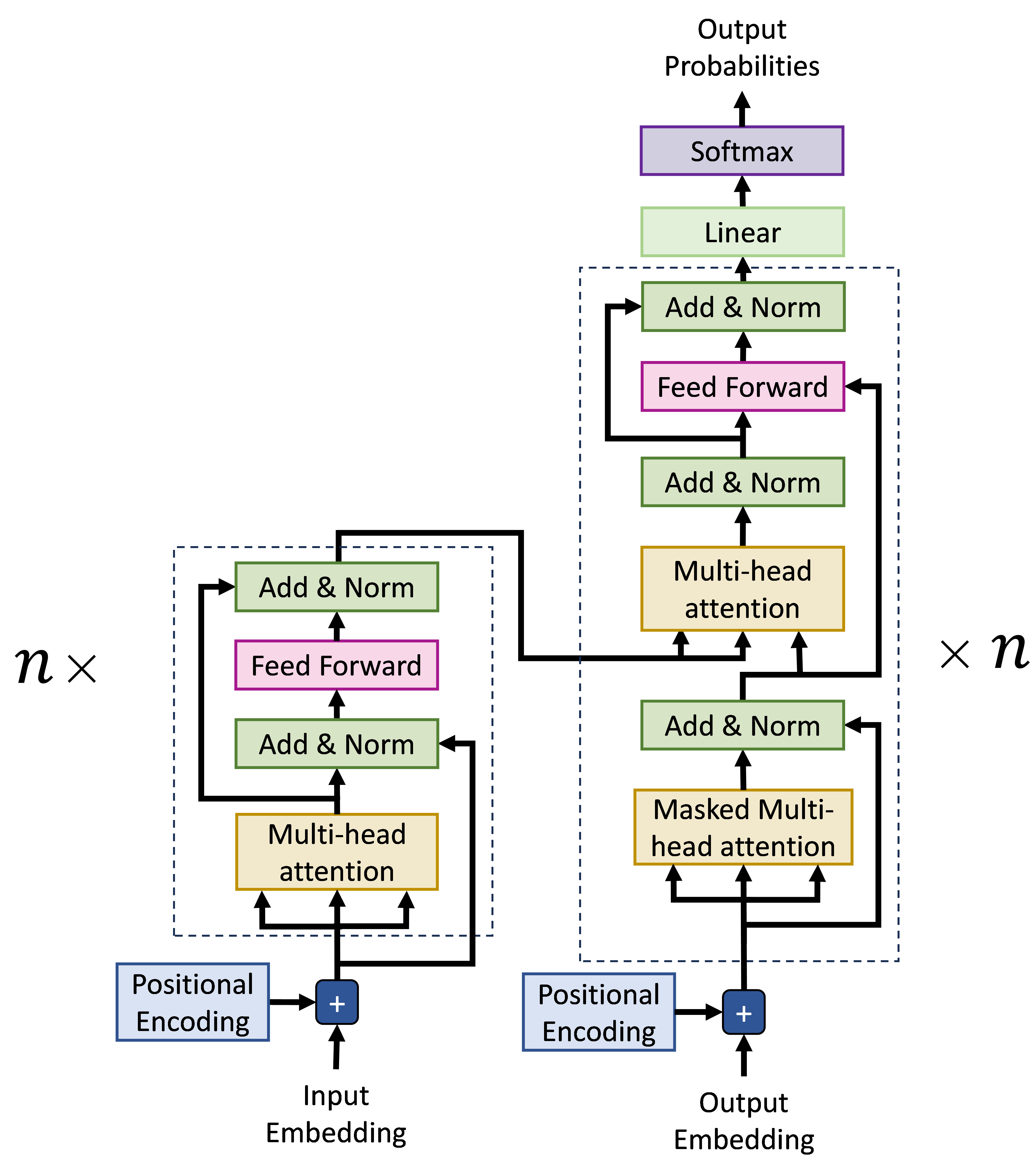}
    \caption{Transformer}
    \label{fig:subfig6}
  \end{subfigure}
  
  \caption{Common deep neural networks used in security and intrusion detection applications.}
  \label{fig:overall_DL}
\end{figure*}
\section{Deep Learning Approaches}
\label{sec: DL_approaches}
This section aims to provide a concise taxonomy of prominent DL approaches and explore their recent related work, specifically in the context of proactive security techniques in {SG}. DL techniques have shown promise in addressing {SG} security challenges. As illustrated in Fig. \ref{fig:overall_DL}, we can identify the methodologies and architectures successfully applied in proactive security measures for {SG}. Additionally, discussing the recent related work allows us to highlight the advancements in DL-based security techniques that have enhanced the resilience and protection of SG infrastructure against cyber threats.
\subsection{Taxonomy of Deep Learning Approaches}
\subsubsection{Autoencoders}
 {DL} autoencoders (AE) are powerful tools for acquiring efficient unlabeled data representations. These architectures consist of an encoder network that compresses input data into a lower-dimensional latent representation and a decoder network that reconstructs the original input from the compressed representation. Bottleneck refers to a network with a hidden layer containing key data characteristics, significantly smaller than input and output layers \cite{DL_AE1, DL_AE2}. The AE can learn to recognize variations from expected patterns by adjusting model weights and minimizing reconstruction errors during training.

\subsubsection{Recurrence Neural Networks}
Recurrence neural networks (RNNs) are specific types of neural networks capable of handling sequential data. These networks are designed to retain information from previous inputs, allowing them to account for context and dependencies between different time steps. LSTM networks and Gated Recurrent Units (GRU) are two examples of the various types of RNNs. LSTMs and GRUs were developed to tackle the ``vanishing gradients'' issue in recurrent neural networks \cite{DL_RNN2}, which arises when the gradients of the network's weights become very small, making it hard for the network to learn. {LSTMs use specialized memory cells with gating mechanisms to control the flow of information, allowing for access to long-term dependencies \cite{DL_RNN1}.}
On the other hand, GRUs use a simpler gating mechanism relying on a single ``update gate'' to regulate the flow of information, which allows for faster training and less memory usage but may not be as effective in capturing long-term dependencies \cite{DL_RNN3}. Nonetheless, they are still capable of retaining long-term dependencies. Compared to LSTMs, GRUs are faster to train and require fewer parameters, making them more efficient for specific tasks. However, the performance of each model variant relies on the particular task and dataset being utilized.

\subsubsection{Convolutional Neural Networks}
Convolutional Neural Networks (CNN) are a powerful type of neural network that has revolutionized the field of computer vision \cite{DL_CNN1}. These networks use a hierarchical structure of layers to automatically learn increasingly complex features from input data, such as images. The layers perform various functions, including convolution, pooling, and classification. 
{Convolutional layers use filters to learn features such as edges and shapes. These features are then passed to pooling layers, which reduce spatial dimensions, speeding up computation and lowering the risk of overfitting. Pooling layers extract crucial features that enhance image recognition and classification performance. Finally, fully connected layers use these refined features for image classification.}

\subsubsection{Deep Reinforcement Learning}

Deep Reinforcement Learning (DRL) is a {ML}strategy that combines deep neural networks and reinforcement learning techniques \cite{DL_DRL1}. This combination enables the agent to learn from its environment through trial and error. DRL is categorized primarily into two groups: value-based and policy-based. Value-based methods in DRL primarily concentrate on acquiring an optimal value function or Q-value function of various states. This function predicts the anticipated cumulative reward the agent can obtain by selecting a specific action in a specific state. Deep Q-learning (DQL) is a prevalent value-based DRL algorithm that utilizes a deep neural network to estimate the Q-value function. It is an extension of Q-learning. Double Deep Q-learning (DDQL) and Dueling Deep Q-learning (DDDQL) are further extensions of DQL that improve its performance by addressing challenges such as reducing the computation required for training and overestimation bias. Conversely, policy-based DRL algorithms focus on directly finding the optimal policy by mapping states to actions using a neural network. Actor-critic, a frequently used DRL algorithm, combines an actor-network that learns the policy and a critic-network that learns the value function \cite{DL_DRL2, DL_DRL3}.
\subsubsection{Transformers}
The transformer network, proposed by Ashish Vaswani et al. \cite{DL_T1} has gained significant success in Natural Language Processing (NLP) applications. It employs multi-head attention and self-attention, addressing the challenge of capturing long-range dependencies in sequential data \cite{DL_T2}. Transformers have been adapted to computer vision domains and have significantly impacted NLP tasks. The transformer architecture includes an encoder and decoder, each playing a distinct role in the network's operations. The encoder layer transforms input sequences into continuous representations, capturing information through multi-headed attention and a fully connected network. The decoder layer generates text sequences. It utilizes two multi-headed attention layers, a pointwise feed-forward layer, and residual connections. Additionally, it includes a linear layer as a classifier and a softmax operation for word probabilities.
Table \ref{tab:DL_table} provides an extensive overview of various DL approaches for cybersecurity applications, including information on their specific approach, suitable use cases, advantages, disadvantages, and relevant research papers.

\subsubsection{Graph Neural Networks}
Graph Neural Networks (GNNs) are DL models that capture intricate relationships and dependencies in data organized in a graph structure. They learn about nodes and edges by collecting information from neighboring nodes and adjusting node features based on this amalgamated information. GNNs use node properties and connections stored in adjacency matrices to generate embeddings that capture both the structure and characteristics of nodes. These embeddings are useful for making predictions at the individual node and graph levels \cite{DL_GNN1}. GNNs have various variants, each with distinct capabilities for modeling and analyzing graph-structured data. Recurrent GNNs (RecGNNs) use recurrent neural architectures to acquire node representations through continuous information exchange. Convolutional GNNs (ConvGNNs) expand upon convolution to incorporate graph data, aggregating nodes, and neighboring features.
\begin{table*}[]
\centering \caption{Overview of DL Approaches for Cybersecurity Applications.} \footnotesize
\begin{tabular} {|p{1.5cm}|p{4cm}|p{4.5cm}|p{4cm}|p{1.5cm}|}
\hline
\textbf{Approach} & \textbf{Where to Use} & \textbf{Pros} & \textbf{Cons} & \textbf{Papers} \\ \hline
Autoencoders & DDoS attack detection, network flow characteristics, intrusion detection, zero-day malware detection, anomaly detection in Battery Electrical Storage Systems, hierarchical detection approach. & High prediction accuracy, effectiveness in identifying anomalies, robust to adversarial attacks, ability to detect anomalies with small variations over time, ability to learn complex correlations between measures. & Accurate estimation of the state of charge remains a limitation. & \cite{AE1, AE2, AE3, AE4, AE5, AE6, AE7} \\ \hline
DRL & Data integrity threats in AC power systems, data integrity breaches, FDI attacks in the SG, data integrity attacks in AC power systems, FDI attacks on dispersed networks, FDI attacks on energy pricing. & High learning efficiency, improve detection performance, identify FDI attacks and engage in more rewarding actions, reduce power losses in the SG. & Continuous attacks have a long-term influence on system SE, making it challenging for detection systems to capture them. &\cite{DRL1, DRL2, DRL3, DRL4, DRL5, DRL7, DRL8, DRL9} \\ \hline
RNN & Two-phase learner-based FDI attacks detection technique, real-time detection mechanism within the physical microgrid, AGC systems SE, DC microgrid vulnerability to FDI attacks on the DC bus voltage, {electricity theft.} & Captures non-linear data features, real-time evaluation, reduces the impact of {FDI attacks}, accurately estimates the attack signal and outperforms classical neural networks in disturbance rejection and tracking performance. & RNN models are complex and process sequential data, leading to training complexity. They have limited interpretability, making explaining predictions or identifying patterns difficult. Long sequences, like time series data in {SG}, can introduce computational and memory limitations, affecting performance and efficiency. & \cite{RNN1, RNN2, LSTM1, LSTM2, GRU1, RNN3, RNN4, RNN5} \\ \hline
CNN & 
Use CNN for detection of energy theft in SG. They are effective against DoS, FDI attacks, jamming, and hybrid attacks. Additionally, CNN can be applied to HVDC systems and detect FDI attacks on Phasor Measurement Unit (PMU) data. & 
CNN efficiently identifies abnormal patterns, learns relevant features, simplifies training, and create deep networks with low computational demands. & CNN lacks sequence modeling capabilities to capture long-term dependencies in sequential data. Handling variable-length sequences also poses challenges for CNNs. &\cite{CNN1, CNN2, CNN3, CNN4, CNN5} \\ \hline

GNNs & Detecting cyber attacks, identifying attacks, identifying cyber vulnerabilities in grid-connected photovoltaic (PV) inverters, detecting FDI attacks on smart power grids. & considering both temporal measurements and topological connectivity, captures the spatial correlations of SG data, integrates the spatial aspects of power system topology into the detection mechanism. & Lacks temporal information, limiting timely response and accuracy.  The model's real-time applicability is constrained, and its capacity for dynamic parameter tuning is limited. Less reliable for scenarios with varying attack locations and intensities. & \cite{GNN1, GNN3, GNN4, GNN5, GNN6} \\ \hline
CNN + LSTM & Identify cyber vulnerabilities in grid-connected {PV} inverters, identify energy theft in AMI systems, identify stronger correlations between Time-varying data. & Minimize the spectral fluctuation across several variables, minimize temporal fluctuation in the data, capture spatial features using the convolutional layer and temporal features using the LSTM. & The complexity of CNN-LSTM models necessitates additional data and resources for training. Tuning hyperparameters becomes a time-consuming task. Interpretability is challenging due to complex intermediate representations. & \cite{Others_LSTM_CNN1, Others_LSTM_CNN2, Others_LSTM_CNN3, Others_LSTM_CNN4, Others_LSTM_CNN5}\\ \hline
Transformers & Anomaly detection of Time Series Data, Advanced Persistent Threat (APT) attacks in IIoT systems, DDoS attacks in SDN, and detecting stealthy FDI attacks in SG. & Transformer-based sequence modeling and forecasting architecture offer parallel efficiency and long-distance context information. It addresses FDI attack detection challenges and captures long-term correlations, APT attacks, and contextual information in IoT systems, enhancing computational efficiency and scalability. & Transformers prioritize global relationships, which may limit their ability to capture local patterns and temporal dependencies effectively. They are computationally expensive, slow, and time-consuming, making them challenging for large-scale time series data and experimentation.
&\cite{Transformners_1, Transformners_2, Transformners_3, Transformners_4, Transformners_5, Transformners_6, Transformners_7, Transformners_8} 
\\ \hline
\end{tabular}
\label{tab:DL_table}
\end{table*}
\subsection{DL-based Prominent Approaches}
\subsubsection{Autoencoders}
{AE models can extract relevant features, making them valuable in detecting anomalies by distinguishing between signal and noise components. This makes them popular for fault and intrusion detection \cite{AE1,AE2}.}
The authors in \cite{AE1} proposed a DDoS attack detection system using multilevel shallow and deep AE. The model employed an unsupervised technique to encode training data for dimensionality reduction and use multiple kernel learning to combine encoded features. This approach improves detection accuracy by automatically considering AE-based characteristics, surpassing LSTM and random forest.

Andresini et al. \cite{AE2} used 1D arrays of network flow characteristics to train intrusion detection models. They assessed the effectiveness of 1D CNNs, AEs, and multi-channel convolutions. They integrated supervised and unsupervised multi-channel feature learning to identify feature dependencies and enhance detection effectiveness by learning the impact of each channel and utilizing existing patterns among channels.

Authors in \cite{AE3} presented ICVAE-DNN for intrusion detection, combining ICVAE with DNN to identify sparse representations in large datasets. The ICVAE encoder sets DNN hidden layer weights, enabling fast training and avoiding local minima. It generates unidentified attack samples to balance the training data and increase variability, allowing DNN to extract high-level features and reduce input dimensionality.

Kim et al. \cite{AE4} introduced an AE-based semi-supervised model and One-Class classification algorithm for zero-day malware detection, combining VAE and CNN AE(1D) models. The method is resilient against evasion and adversarial attacks, demonstrating the potential for identifying new and unseen malware without manual configuration.

The authors in \cite{AE5} propose an anomaly detection algorithm for Battery Electrical Storage Systems in DERs. The algorithm utilizes an AE to detect dangerous working conditions by observing the physical behavior of the system. It outperforms traditional One-Class Support Vector Machines in detecting unsafe conditions, unusual behavior, partial attacks, and bad data injection. However, accuracy in estimating charge remains a limitation.

Kye et al.~\cite{AE6} introduced a hierarchical detection approach using multiple stages to identify extreme anomalies based on abnormality levels. The NIDS solution improves detection performance without relying on scarce abnormal data. It uses the autoencoder's hidden space, specialized anomaly scores, and self-supervision signals during training. The detection process evaluates anomaly scores and triggers alarms based on a predefined threshold. The first stage identifies extreme outliers using meta-representation, while the second stage amplifies the reconstruction error and detects anomalies missed in the first stage~\cite{AE6}.
{Song et al. \cite{AE7} investigated network intrusion detection using AE models and found that model capacity significantly impacts effectiveness. Models with larger capacities performed better, but the number of hidden layers did not correlate substantially with its efficacy, suggesting the need for a suitable model structure and latent size.

}

\subsubsection{DRL}
{The Deep-Q-Network Detection (DQND) algorithm has been a crucial technique utilized in various studies to enhance cybersecurity in power systems. Authors in \cite{DRL1} applied the DQND technique to improve the defensive approach against data integrity threats in AC power systems SE. The effectiveness of the learning process was investigated using a quantification of the observation space sliding window.} {Meanwhile, Lee et al. \cite{DRL7} developed a DQND algorithm to reduce detection delays in cyberattacks on SG, ensuring rapid and precise threat identification.}
Authors in \cite{DRL2} proposed a DRL-based method for detecting data integrity breaches using an LSTM network to assess power grid attacks. The weights are subjected to noise for agent exploration, and a multi-step reward function improves Q-value accuracy. Continuous attacks have a long-term impact on system SE, making detection challenging compared to discontinuous attacks, which can only be detected at discrete time steps.

Huang et al. \cite{DRL3} proposed an attention-based DRL detection model to identify FDI attacks in the SG. The model uses an attention mechanism to extract state information, assigning weights to state features and learning the significance of each measurement point. {This enables the agent to make informed decisions by focusing on significant pre- and post-attack state features.} 
Using multi-sourced data, Hu et al. \cite{DRL5} adaptive feature boosting method for intrusion detection in {SG}. The technique employs weighted feature sampling and a series of  AE to capture critical features and train them to a random forest classifier. The Deep Deterministic Policy Gradient (DDPG) determines feature sampling probability. The ensembled AE and classifier sequence differentiate between normal, fault, and attack events.
Abianeh et al. \cite{DRL8} developed a DDPG multiagent DRL to analyze index-based detection algorithm vulnerabilities against stealthy destabilizing {FDI attacks} on dispersed networks and launch coordinated disruptive FDI attacks on cyber-secured DC microgrids. The algorithm integrated a sniffer feature to detect and resolve disconnections but failed to detect the coordinated FDI attacks. An additional RL DQN detection algorithm was presented to improve the reliability of index-based {FDI attacks} susceptibilities detection algorithms, yielding a fully cyber-secure system.
Zhang et al. \cite{DRL9} developed a two-stage defensive approach to address security risks in {FDI attacks} on energy pricing. They combined recursive least-square methods with a radial basis function neural network to improve learning efficiency. The distributed DDPG approach splits actor networks into multiple subproblems, enabling microgrids to learn their best action schemes independently while collaborating with nearby microgrids.
Similarly, Selim et al. \cite{DRL4} proposed an approach to address voltage violations and reduce power losses in the SG by regulating DERs and tie-switches. The original soft actor-critic (SAC) technique has been expanded to accommodate both discrete and continuous actions for scenarios, including regulating DERs' setpoints and network switches.

\begin{table*}[]
    \centering    \caption{{DL} countermeasures for cybersecurity in power systems.} {\footnotesize
    \begin{tabular}{|c|p{4.5cm}|c|cccc|ccc|c|c|c|}
    \hline
    \multirow{2}{*}{Paper} & \multirow{2}{*}{Security Issue} & AE & \multicolumn{4}{c|}{DRL} & \multicolumn{3}{c|}{RNN} & CNN & GNNs & Transformers \\
    \cline{4-10}
     & & & DQN & DDQN & SAC & DDPG & RNN & LSTM & GRU & & & \\
    \hline
    \cite{AE1} & DDoS attack detection & \cmark & & & & & & & & & & \\  \hline
    \cite{AE2} & Intrusion detection & \cmark & & & & & & & & \cmark & & \\  \hline
    \cite{AE3} & Intrusion detection &  \cmark & & & & & & & & & & \\  \hline
    \cite{AE4} & Zero-day malware detection & \cmark & & & & & & & & \cmark & & \\  \hline
    \cite{AE5} & Anomaly detection for battery systems & \cmark & & & & & & & & & & \\  \hline
    \cite{AE6} & Extreme anomalies detection &  \cmark & & & & & & & & & & \\  \hline
    \cite{AE7} &  {AE for network intrusion detection }&  \cmark & & & & & & & & & & \\  \hline
    \cite{DRL1} & Data integrity in AC power systems & & \cmark & & & & & & & & & \\  \hline
    \cite{DRL2} & Data integrity breaches & & \cmark & & & & & \cmark & & & & \\  \hline
    \cite{DRL3} & False Data Injection Attacks (FDI attacks) & & & \cmark & & & & & & & & \\  \hline
    \cite{DRL4} & Intrusion detection in {SG} & & & & \cmark & & & & & & & \\  \hline
    \cite{DRL5} & Data integrity attacks in AC power system & \cmark & & & & \cmark & & & & & & \\  \hline
    \cite{DRL7} & {FDI attacks} on dispersed networks & & & \cmark & & & & & & & & \\  \hline
    \cite{DRL8} & {FDI attacks} on energy pricing  & & & & & \cmark & & & & & & \\  \hline
    \cite{DRL9} & {FDI attacks} on energy pricing  & & & & & \cmark & & & & & & \\  \hline
    \cite{RNN1} & FDI attacks detection & & & & & & & \cmark & & & & \\  \hline
    \cite{RNN2} & {FDI attacks} on smart microgrids & & & & & & \cmark & & & & & \\  \hline
    \cite{LSTM1} & Impact of FDI attacks and TD attacks on AGC & & & & & & & \cmark & & & & \\  \hline
    \cite{LSTM2} & TD attacks detection & & & & & & & \cmark & & & & \\  \hline
    \cite{GRU1} & FDI attacks on the DC bus voltage & & & & & & & & \cmark & & & \\  \hline
    \cite{RNN4} & {Electricity theft in SG} & & & & & & \cmark & & \cmark & & & \\  \hline
    \cite{RNN5} & {Electricity theft in SG} & & & & & & & \cmark & \cmark & & & \\  \hline
    \cite{RNN3} & {SDN-based defense against SG cyber-attacks} & & & & & & & & \cmark & & & \\  \hline
    \cite{Others_LSTM_CNN1} & Cyber vulnerabilities in grid-connected PV inverters 
    & & & & & & & \cmark & & \cmark & & \\  \hline
    \cite{Others_LSTM_CNN2} & Energy theft in AMI systems & & & & & & & \cmark & & \cmark & & \\  \hline
    \cite{Others_LSTM_CNN3} & ETD & & & & & & & \cmark & & \cmark & & \\  \hline
    \cite{Others_LSTM_CNN4} & Electricity theft in {
SG} & & & & & & & \cmark & & \cmark & & \\  \hline
    \cite{Others_LSTM_CNN5} & Vulnerability of SCADA systems to FDI attacks & & & & & & & \cmark & & \cmark & & \\  \hline
    \cite{Others_GRU_CRNN} & Cyber-attacks affecting energy generation readings & & & & & & & & \cmark & \cmark & & \\  \hline
    \cite{CNN1} & Cyber-attacks on PMU-based HVDC Ancillary Service Control 
    & & & & & & & & \cmark & \cmark & & \\  \hline
    \cite{CNN4} & Multiple cyber-attacks on Cyber-Physical Power Systems 
    & \cmark & & & & & & & & \cmark & & \\  \hline
    \cite{CNN2} & Energy theft detection & & & & & & & & & \cmark & & \\  \hline
    \cite{CNN3} & Energy theft detection & & & & & & & & & \cmark & & \\  \hline
    \cite{CNN5} & Privacy-preserving electricity theft detection & & & & & & & & & \cmark & & \\  \hline
    \cite{GNN1} & Attack identification and localization & & & & & & & & \cmark & & \cmark & \\  \hline
    \cite{GNN3} & FDI attacks detection & & & & & & & & & \cmark & \cmark & \\  \hline
    \cite{GNN4} & Cyberattack identification & & & & & & & & & & \cmark & \\  \hline
    \cite{GNN5} & Large-scale AC power grid FDI attacks detection & & & & & & & & & & \cmark & \\  \hline
    \cite{GNN6} & FDI attacks detection on smart power grids & & & & & & & & & \cmark & \cmark & \\  \hline
    \cite{Transformners_1} & APT attack detection & & & & & & & & & &  & \cmark \\  \hline
    \cite{Transformners_2} & IoT anomaly detection & & & & & & & & & & \cmark & \cmark \\  \hline
    \cite{Transformners_3} & DDoS attack detection on SDN & & & & & & & & & \cmark &  & \cmark \\  \hline
    \cite{Transformners_4} & FDI attacks detection in {SG} & & & & & & & & & &  & \cmark \\  \hline
    \cite{Transformners_5} & IDS for high-dimensional anomaly data & & & & & & & & & &  & \cmark \\  \hline
    \cite{Transformners_6} & Anomaly detection on time series data & & & & & & & & & &  & \cmark \\  \hline
    \cite{Transformners_7} & Time series anomaly detection & & & & & & & & & \cmark &  & \cmark \\  \hline
    \cite{Transformners_8} & Intrusion detection & & & & & & & & & &  & \cmark \\  \hline
    \end{tabular}}
\label{tab:DL_CyberSecurity}
\end{table*}

\subsubsection{RNN}
\label{RNN_LSTM}
Wang et al.~\cite{RNN1} introduced the KFRNN system, a two-phase learner-based FDI attacks detection technique. The first phase utilizes RNN to capture non-linear data features and the Kalman filter for linear data features. The second-phase learner dynamically merges the outcomes of the two initial learners using a fully connected layer and backpropagation module. Results show that the KFRNN approach outperformed the conventional Bad Data Detector and R2N2\_variant.
Naderi et al. \cite{RNN2} proposed a remedial action scheme to combat severe {FDI attacks} targeting smart microgrids. The scheme uses a hardware-in-the-loop integrated with an RNN for real-time detection within the physical microgrid. The model includes six parallel RNN units, two for each device type. Microgrid sensors provide inputs, including voltage and current readings injected with FDI attacks. The RNN units use identical layer structures for predictions, estimating new readings, and comparing them to sensors' values.
Authors in \cite{LSTM1} explored the impact of nonlinearities on AGC's response to FDI attacks and Time Delay (TD) attacks, highlighting the need to consider these factors during AGC system design and implementation. The model consists of two stages: detecting and identifying attacks using an LSTM model trained on a normal AGC operation dataset, and reducing the impact of {FDI attacks} using another LSTM model. This technique uses compromised signals from the first step to produce a rectified signal and employs Root Mean Square Error (RMSE) to assess its effectiveness.
Ganesh et al. \cite{LSTM2} developed a hierarchical LSTM model to monitor and detect TD attacks online, processing raw data streams from sensors. The model consists of two layers linked sequentially, segmenting input data into shorter sub-sequences and extracting local properties independently. The upper level evaluates local features to obtain temporal properties for the entire sequence. This hierarchical structure captures both short-term and long-term dependencies, enhancing its effectiveness in recognizing TD attacks.
In \cite{GRU1}, the authors proposed a GRU-based framework to address DC microgrid vulnerability to FDI attacks on the DC bus voltage. The framework consists of two parts: an offline-trained GRU network for real-time evaluation of the DC bus voltage, which serves as the estimation strategy, and a GRU network combined with a PI controller to counteract {FDI attacks}, known as the mitigation strategy. 

Authors have investigated the integration of various RNN variants and other DL techniques to tackle different cyber-attacks in {SG}, including the combination of CNN and LSTM methods \cite{Others_LSTM_CNN1, Others_LSTM_CNN2, Others_LSTM_CNN3}.
Mao et al. \cite{Others_LSTM_CNN1} developed a CNN-LSTM technique to identify cyber vulnerabilities in grid-connected {PV} inverters, reducing cybersecurity threats from the growing use of renewable energy in power systems. CNN layers minimize spectral fluctuation across variables, allowing flexible interpretation after reshaping, while LSTM layers minimize temporal fluctuations. Input de-redundancy and hyperparameter selection maximize detection.
Adhure et al. \cite{Others_LSTM_CNN2} designed an Electricity Theft Detection (ETD) model for identifying energy theft in AMI systems using a CNN-LSTM stack. The model utilizes one-dimensional convolutional layer data as input with a kernel size of five and five filters. It includes two LSTM layers (48 and 24 neurons) with 20\% dropout layers.
The study in \cite{Others_LSTM_CNN3} introduces a ConvLSTM-based model for ETD, incorporating a reshaped 2D matrix of consumption data and ConvLSTM with embedded CNN and batch normalization. The model replaces the pooling layer with LSTM to identify stronger correlations in time-varying data, capturing spatial and temporal features. It improves dataset balancing, captures local electricity usage, and uses batch normalization for efficient deployment.
The work proposed in \cite{Others_LSTM_CNN4} aims to combat electricity theft in {SG} by installing central observer meters in communities. The technique analyzes users' electricity consumption patterns by transforming time series data into a structured representation, considering temporal elements and variations across neighboring periods and days. This approach addresses localized theft concerns and enhances the efficiency of {SG}.
In \cite{Others_LSTM_CNN5}, a two-stage framework is introduced to tackle the vulnerability of SCADA systems in {SG} to FDI attacks. It combines real-time data forecasting through a CNN-LSTM autoencoder sequence-to-sequence architecture with anomaly detection based on the forecast. The model accurately estimates the power consumption state and incorporates an adaptive threshold to identify abnormal behavior. 
Ismail et al. \cite{Others_GRU_CRNN} developed an ETD system using deep neural networks to identify cyber-attacks affecting energy generation readings. They explored three architectures: deep feed forward, deep recurrent, and deep convolutional-recurrent neural networks. The deep feed-forward network processed energy generation readings, while the deep recurrent network with GRUs captured temporal correlations, processed data, and predicted class labels. The convolutional-recurrent neural network (C-RNN) hybrid architecture combined convolutional and max-pooling layers for feature extraction and class label prediction.
{Electricity theft in SG using a GRU-RNN model was optimized for time-series data utilizing month and weekday features, focusing on small perturbation attacks and showing superior performance even with minor energy reporting modifications \cite{RNN4}. While \cite{RNN5} uses GRU and Bidirectional LSTM for detecting theft in solar and wind in distributed generation units, with GRU for single fuel types and BLSTM for complex hybrid fuel scenarios.
Kumar et al. \cite{RNN3} developed a model that integrates a blockchain-based authentication scheme with a {DL}-based IDS within an SDN framework to improve cybersecurity in SG networks. The model uses a self-attention mechanism and a Bi-Gated Recurrent Unit for temporal dependencies classification.}

\subsubsection{CNN}
Authors in \cite{CNN1} introduced a SE-DCNN variant to improve PMU-based HVDC Ancillary Service Control (HASC) reliability in cyber-attack-prone environments. This CNN structure combines time and frequency domain information for feature extraction, addressing redundant features. A Squeeze-Excitation structure addressed redundant features by assigning weights to useful ones and reducing redundant ones. The SE-DCNN model explicitly models channel interdependencies, enhancing sensitivity to informative attack characteristics and enabling more accurate and secure detection.
The RL-CNN approach focuses on intelligent attack localization and recovery in Cyber-Physical Power Systems under multiple cyber-attacks \cite{CNN4}. It involves a complex system consisting of a representation-learning-based DL model, deep AE, and a CNN architecture. The detection problem involves multilabel classification, and system recovery is performed based on attack information. The MMSE estimator minimizes mean-squared error and provides conditional expectations for the system state.

Several authors addressed the problem of energy theft detection in the SG \cite{CNN2, CNN3, CNN5}. Yao et al. \cite{CNN2} proposed an approach that combines privacy preservation and DL techniques. They utilize Paillier Homomorphic Encryption to protect energy usage privacy and employ a combined CNN model for analyzing decrypted SM data. The model classifies the data as either theft or normal. It employs preprocessing techniques to format the data into image-like structures, allowing the CNN model to extract relevant features for accurate classification. 
In \cite{CNN3}, the authors present a novel ETD approach using a large dataset of energy usage records from smart meters. They use a deep CNN model with convolution and pooling layers for feature extraction and classification. This approach emphasizes analyzing power usage patterns to differentiate between normal consumers and potential energy thieves, addressing data augmentation challenges. 
The p2Detect framework improves privacy-preserving ETD by optimizing model efficiency and accuracy through homomorphic encryption-friendly training, addressing limitations in approximations and clustering~\cite{CNN5}. It employs secure cryptographic protocols, ciphertext blinding, and padding embedding to safeguard data and model privacy. However, it fails to consider security threats from dynamic cyberattacks. Integrating online learning and using captured samples for incremental learning is recommended. Secure inferencing using encrypted neural networks faces declining efficiency due to growing homomorphic encryption parameters~\cite{CNN5}.

\subsubsection{GNN}
\label{GNN}

Since the power system can be modeled as a graph, with buses represented as vertices and transmission lines as edges, many researchers have explored using GNNs to detect cyber attacks \cite{GNN1,GNN3}.
Haghshenas et al. \cite{GNN1} proposed a TGNN framework that considers temporal measurements and topological connectivity to identify attacks. The model uses GNNs with a message-passing technique, a GRU for temporal data, and a residual block for vanishing gradients. The effectiveness is assessed using ramp attack and FDI attack data, revealing high accuracy in identifying and locating attacks. The study also evaluates the model's sensitivity to attack location and severity, revealing areas where attacks are harder to detect.
Similarly, authors in \cite{GNN3} proposed a scalable and real-time FDI attack detector as an early detection and identification system.
Ruan et al. \cite{GNN4} proposed a strategy using spatiotemporal graph {DL} (STGDL) to identify cyberattacks. The approach extracts spatiotemporal features using graph convolution and temporal gated convolution. Quantile regression training is utilized to create safe boundaries for state variables in SE to identify abnormalities caused by cyberattacks. The proposed super-resolution perception network improves temporal learning by creating high-frequency data using low-frequency SE results. 
Boyaci et al. \cite{GNN6} proposed a Chebyshev Graph Convolutional Networks (CGCN) detector to capture the spatial correlations of SG data and efficiently model the graph structure of large-scale AC power grids addressing {FDI attacks}. The suggested architecture's autonomous training method eliminates the need for any optimization steps, making it more suitable for practical applications. 
The work in \cite{GNN5} proposed a GNN-based FDI attack detection strategy on smart power grids using a topology-aware GAE. It addresses the shortcomings of existing detectors, including the lack of spatial embedding or detection of only types of {FDI attacks} found in the training set. The strategy integrates power system topology spatial aspects to better detect unseen {FDI attacks} enhancing generalization abilities. Table \ref{tab:DL_CyberSecurity} summarizes the existing works on {DL} countermeasures for enhancing cybersecurity in power systems.

\subsubsection{Transformers}
The transformer {DL} technique has been widely used for anomaly detection.
BERT is a proactive APT detection scheme for IIoT systems, overcoming limitations in traditional representations like word2vec~\cite{Transformners_1}. It models global dependencies and encodes the APT attack sequence dataset using BERT's bidirectional transformer encoder. A Softmax regression classifier trains the model for accurate classification, leveraging BERT's transformer architecture and contextual understanding.
The GTA framework, presented in \cite{Transformners_2}, employs a transformer-based architecture for IoT anomaly detection. It incorporates a connection learning policy using Gumbel-softmax sampling to automatically learn the graph structure, revealing hidden associations for accurate forecasting. The framework also limits the neighborhood scope of each node, enabling efficient anomaly detection.
An Influence Propagation convolution is introduced to capture anomaly information flow.
Wang et al. \cite{Transformners_3} proposed DDosTC, a novel hybrid neural network model designed to detect DDoS attacks on SDN. By integrating scalable transformers and CNN  architecture, this model effectively addresses vulnerabilities and mitigates the risk of potential network collapse. DDosTC comprises a transformer layer, CNN layer, and dense layer for feature extraction.
Li et al. \cite{Transformners_4} proposed a method for detecting {FDI attacks} in {SG} using transformer, FL, and the Paillier cryptosystem. Distributed detection with edge node detectors eliminates central workstation and communication delays. The transformer model analyzes electrical quantities at edge nodes, allowing precise FDI attack detection. The FL and the Paillier cryptosystem encrypt exchanged data, enhancing security and ensuring confidentiality against hackers.
The Robust Transformer-based Intrusion Detection System (RTIDS) is an efficient and accurate IDS for high-dimensional anomaly data \cite{Transformners_5}. It uses techniques like positional embedding, stacked encoder-decoder neural networks, and self-attention mechanisms to reconstruct feature representations and classify network traffic. The framework consists of three modules: data preparation, RTIDS model construction, and real-time detection.
The authors in \cite{Transformners_6} proposed TGAN-AD to address the challenges in anomaly detection on time series data. It combines Generative Adversarial Networks (GANs) and transformers to extract contextual features from time series data and employs a discriminator to identify abnormal data. TGAN-AD demonstrates high accuracy in anomaly detection across multiple datasets, such as SWaT, WADI, and KDDCup99.
DCT-GAN is a novel approach for time series anomaly detection, overcoming challenges like model collapse, low generalization capability, and poor accuracy \cite{Transformners_7}. It employs multiple generators and a single discriminator, enabling the extraction of fine-grained and coarse-grained information from time series data. It also incorporates a weight-based mechanism to balance the contributions of different generators. 
The authors in \cite{Transformners_8} proposed RUIDS, a robust unsupervised intrusion detection system incorporating a masked context reconstruction module into a transformer-based self-supervised learning scheme. It aims to capture temporal context and mitigate anomaly contamination in network intrusion data. The scheme learns intrinsic relationships within contexts using learnable transformations and a transformer module, improving system robustness and showing promise in enhancing intrusion detection using novel techniques.

\subsection{MTD DL-based Approaches}
\label{sec: MTD & DL}
Integrating MTD and DL techniques in {SG} offers a promising enhanced security and robustness strategy. MTD dynamically changes system configurations, while DL allows {SG} to learn and detect anomalous patterns adaptively, enabling proactive threat mitigation and rapid response. This synergy enhances SG security, protects critical infrastructure, preserves data integrity, ensures privacy, and maintains an uninterrupted energy supply.

Several approaches included using MTD to enhance the robustness of DL models against adversarial attacks. The work of \cite{MTD_DL_2}, aims to address the susceptibility of DL-based visual sensing systems to adversarial example attacks using MTD strategy. The proposed approach deploys multiple models that work collaboratively to detect adversarial example attacks and then use MTD to generate new models. Serial data fusion with early stopping is also employed to improve defense effectiveness and efficiency by reducing inference time and enhancing real-time performance. 
Morphence employs adversarial training on a subset of the model pool \cite{MTD_DL_3}. It increases the number of models to challenge adversaries through weight perturbation, data transformations, and integration. A scheduling strategy selects the most confident model for predictions, and the model pool is automatically renewed within a predetermined query budget. It also provides an effective defense against repeated and correlated attacks.
Li et al. \cite{MTD_DL_6} introduced the wAdvMTD approach, using pre-built models like 3-layer CNN and ResNet18 for classification. The mechanism generates functionally equivalent but structurally different models, making the attack vector ineffective against most models, regardless of adversary knowledge of model selection. It consists of training and discrimination stages, with multiple candidate models working together and using differential voting to mitigate perturbations and ensure system resilience against single-point failures.
Xu et al. \cite{MTD_DL_4} proposed an approach to counter stealthy FDI attacks in power system SE. The approach uses LSTM-AE to analyze and learn spatiotemporal correlations among adjacent measurements, improving detection accuracy and controlling false alarms. It uses iterative normality projection to identify attack vector parts. The MTD algorithm uses D-FACTS devices to inject unknown reactance into the grid and verify detector decisions. 
Xu et al. \cite{MTD_DL_5} developed a framework combining a data-driven detector and physics-based MTD. The framework uses a single LSTM-AE DL model and power system physics data for accuracy. If a positive alarm is raised, a neural network-based attack detection technique is applied, and a robust MTD algorithm is initiated to confirm the alert. It improves detection performance and achieves a high True Positive Rate while minimizing false positive rates and rejecting false alarms in real-time.

\section{Benchmark Datasets}
\label{sec: Benchmark Datasets}
The selected datasets below provide a distinct range of diverse attacks commonly used for IDS research. While some may fail to include current traffic patterns and emerging attacks, others have been generated recently to address these limitations. We will also discuss the limitations associated with these datasets.
\subsection{KDDCUP99}
The KDDCUP99 dataset is widely used for evaluating network intrusion detection methods. It contains network traffic data with both normal and attack instances, categorized into four types: DoS, User to Root (U2R), Remote to Local (R2L), and Probing \cite{datasets_1}. Each category represents a distinct attack scenario, and 41 features are available for each connection record. This dataset is a benchmark for intrusion detection research, facilitating supervised learning and algorithm evaluation. However, it is essential to acknowledge that the dataset is synthesized based on DARPA'98 data, which introduces concerns about its representation of real network traffic. The dataset's coverage of attack types and variations may be limited, and its distribution of attack types is imbalanced, which can impact its ability to generalize to new patterns. The presence of duplicate records further distorts the evaluation of detection algorithms, potentially yielding biased results and restricted generalization capabilities \cite{datasets_2}. These characteristics highlight the non-stationary nature of the dataset and emphasize the need to account for such divergence to improve the performance of intrusion detection techniques \cite{datasets_3}.

\begin{table*}[]
\centering
\caption{{Comprehensive Analysis of Cyber Defense Strategies in Smart Grids}}
\label{tab:DL_details}

\begin{tabular}{|p{0.7cm}|p{1.5cm}|p{1.5cm}|p{1.5cm}|p{1.7cm}|p{1.8cm}|p{2.5cm}|p{3.2cm}|}
\hline
\textbf{paper} & \textbf{Compromised Security Objectives} & \textbf{Attack Type} & \textbf{Features} & \textbf{Level} & \textbf{Dataset} & \textbf{Results} & \textbf{Limitations} \\ \hline
\cite{AE3} & Availability \newline Integrity \newline Confidentiality & DoS, U2R, R2L, Worms, Shellcode & Network traffic: protocol type, service type, flag status, port numbers, traffic stats: packet count, size, inter-arrival times, etc. & Network & UNSW-NB15, KDDTest, KDDTest-21 & KDDTest+ dataset: \newline Recall: 0.7743, \newline Precision: 0.9739, \newline FPR: 2.74\%, \newline F1-score: 86.27\%, \newline Accuracy: 85.97\%. \newline  UNSW-NB15 dataset: 
\newline  Recall: 0.9568, \newline Precision: 0.8605, \newline FPR: 19.01\%, \newline F1-score: 90.61\%,\newline Accuracy: 89.08\%. & Possible deviation of generated attacks from actual network behavior, KL-vanishing phenomenon might impact the quality of the generated samples, model performance might be sensitive to the training sample size,  attack class reconstruction loss must be carefully calculated for new sample generation.   \\ \hline
\cite{AE5} & Availability \newline Integrity & Improper working condition, Data manipulation, Attacks on SCADA system & Battery state of charge, terminal voltage, cell current, DC link voltage, and AC side phase voltages and currents, etc. & End-User, Control Systems, Data Centers/Cloud & – & The proposed autoencoder outperforms OCSVM swiftly and accurately detects attacks. & Limitations, especially in detecting sophisticated attacks on the State of Charge measure. \\ \hline
\cite{AE2} & Availability \newline Integrity \newline Confidentiality & Network-based cyber-attacks: DDoS, Malware, Phishing, MitM, etc. & Various types of network traffic & End-User, Networks, Control Systems, Data Centers/Cloud & KDDCUP991, UNSW-NB15, CICIDS2017 & KDDCUP991: \newline Accuracy: 92.49\%, \newline F1-score: 95.13\%. \newline UNSW-NB15: \newline Accuracy: 93.4\%, \newline F1-score: 95.29\%. \newline CICIDS2017: \newline Accuracy: 97.9\%, \newline F-score: 94.93\%. & Model's performance varies with dataset characteristics; outperforms on UNSW-NB15Test but second to DNN 4 Layers on KDDCUP99Test, indicating potential dataset-specific optimization needs. \\ \hline
\cite{AE6} &  Integrity & Network Intrusion Detection & Network traffic data & LAN Communication Networks & NSL-KDD, CSE-CIC-IDS2018 & NSL-KDD: \newline Recall: 0.9927, \newline Precision: 0.9648, \newline F1-score: 96.14\%, \newline Accuracy: 98.71\%.\newline CSE-CIC-IDS2018: \newline Recall: 0.9999, \newline Precision: 0.9994, \newline F1-score: 99.98\%, \newline Accuracy: 99.99\%. & Additional delay at the final reexamination stage, especially with deeper autoencoders. \\ \hline
\cite{DRL8} & Integrity & FDI attacks & - & Control Systems & - & Detects stealthy FDI attacks within 2-3s, maintaining low discordant terms, outperforming the discordant method & Potential vulnerability to novel attack vectors not covered in training, dependency on the chosen timestep for detection, and the need to maintain low discordant terms which may be complex in dynamically changing network conditions. \\ \hline
\cite{LSTM2} & Integrity & TD attack & PPCS: pressure, temperature, and electricity generation AGC: tie-line flow measurements & Control Systems & Simulated datasets using Thermopower for PPCS and PowerWorld AGC. & 
Best performing method is 2 HLSTMs+Multi-Task:\newline 
PPCS:\newline MAE: 2.02, \newline  RMSE: 5.53, \newline 
Accuracy: 93.03\%. \newline  AGC:  \newline MAE: 0.47, \newline  RMSE: 0.91, \newline Accuracy: 99.09\%. & Optimal model performance dependent on the correct choice of n and \(\beta\) parameters; FNs can be high for low \(\tau\) values; complexity in balancing error rate and reaction latency. \\ \hline
\cite{GRU1} & Availability & FDI attacks, DC bias injection and time-varying attacks & DC bus voltage, converter output voltage and current & End-user Device and DC microgrid Control Systems
& SGCC Time-series data of output voltage and current from converters within a DC microgrid. & Surpasses classical, Kalman, H-infinity, and hybrid methods in stability, attack mitigation, and disturbance rejection & - \\ \hline
\end{tabular}
\end{table*}

\begin{table*}[]
\centering
\caption*{{Table \ref{tab:DL_details}: Comprehensive Analysis of Cyber Defense Strategies in Smart Grids}} 

\begin{tabular}{|p{0.7cm}|p{1.5cm}|p{1.5cm}|p{1.5cm}|p{1.7cm}|p{1.8cm}|p{2.5cm}|p{3.2cm}|}
\hline
\textbf{paper} & \textbf{Compromised Security Objectives} & \textbf{Attack Type} & \textbf{Features} & \textbf{Level} & \textbf{Dataset} & \textbf{Results} & \textbf{Limitations} \\ \hline
\cite{CNN2} & Integrity \newline Confidentiality & Energy theft attacks & Periodicity and group similarity of energy usage & End-user device and control systems &  SGCC & Accuracy: 92.67\% & - \\ \hline
\cite{CNN4} & Availability \newline Integrity \newline Confidentiality  & FDI attacks, DoS, Jamming, Hybrid & Grid topology data, voltage magnitudes, phase angles, bus measurements. & Communication networks and control systems & Simulation results of IEEE bus systems & 14-Bus  Accuracy: \newline DoS: 96.90\%, \newline FDI: 93.30\%, \newline Jamming: 98.40\%, \newline Hybrid: 98.40\%. \newline 30-Bus Accuracy: \newline DoS: 95.80\%, \newline FDI: 90.50\%, \newline Jamming: 92.10\%, \newline Hybrid: 94.80\%.
& - \\ \hline
\cite{CNN5} & Integrity & Energy theft attacks & Time series electricity consumption data. & Control systems and end-user device & SGCC & Optimized Model \newline AUC: 0.7682. \newline F1-score Range: 28.2\% to 30.9\% (depends on training ratio) & Reduced model complexity for secure inferencing may affect dynamic attack detection; further real-world testing is needed. \\ \hline
\cite{GNN1} & Integrity & FDI attacks, Ramp attacks & Real and reactive power injections, bus voltage, injected bus current, frequency, power injection as time-varying signal. & End-user device, communication networks and control systems & MATPOWER 7.0, NYISO, PSSE & 
FDI attack: \newline Accuracy: 99.50\%, \newline F1-score: 100\%, \newline False Alarm Rate: 0.00\%. \newline Ramp attack: \newline Accuracy: 88.10-89.91\%, \newline F1-score: 99.98\%, \newline False Alarm Rate: 0.02\%. & Sensitivity to attack location, lower detection accuracy for certain buses, some delay in ramp attack detection. \\ \hline
\cite{GNN3} & Availability \newline Integrity \newline Confidentiality & Stealth FDI attacks & Active and reactive power bus injections, and active and reactive power flows on branches obtained from remote terminal units. &  Communication networks and control systems & Synthetic datasets for IEEE bus systems & IEEE 14 F1-score: 90.14\% \newline IEEE 118 F1-score: 94.07\% \newline IEEE 300 F1-score: 97.67\% & - \\ \hline
\cite{GNN4} & Integrity & FDI attacks & Bus voltage, active and reactive powers. & Control systems & Realistic electric load data from the California Independent System Operator 2019 & SRP-STGDL (quantile) excels, surpassing MLP, CNN, LSTM, GRU. Achieved perfect scores on IEEE 30-bus and IEEE 118-bus benchmarks. & - \\ \hline
\cite{GNN6} & Integrity & FDI attacks & Power measurements: voltage, current, and power flow information. & Utility and control systems & Synthetic dataset generated for a 2848-bus test system with 36,000 samples, balanced for attacked/non-attacked data & Detection Rate: 95.53\%, \newline False Alarm Rate: 0.33\%, \newline CGCN Detection Time: 3.25 ms, outperforming FCN, RNN, CNN. & No access to real-world datasets due to privacy reasons, possible discrepancies between synthetic and real-world data. \\ \hline
\cite{Others_LSTM_CNN2} & Integrity & Energy theft attacks & Power consumption patterns. & Control systems & Irish CER & MAE: 0.04541\newline MSE: 0.01426 & Increased computational complexity from CNN-LSTM. \\ \hline
\end{tabular}
\end{table*}

\begin{table*}[]
\centering
\caption*{{Table \ref{tab:DL_details}: Comprehensive Analysis of Cyber Defense Strategies in Smart Grids}} 

\begin{tabular}{|p{0.7cm}|p{1.5cm}|p{1.5cm}|p{1.5cm}|p{1.7cm}|p{1.8cm}|p{2.5cm}|p{3.2cm}|}
\hline
\textbf{paper} & \textbf{Compromised Security Objectives} & \textbf{Attack Type} & \textbf{Features} & \textbf{Level} & \textbf{Dataset} & \textbf{Results} & \textbf{Limitations} \\ \hline
\cite{Transformners_3} & Availability & DDoS & IP flow data, excluding source and destination IPs, ports, timestamps, and flow identification, focusing instead on packet characteristics & Control systems & CICDDoS2019 & Accuracy: 99.70\%, \newline Precision: 0.9998, \newline F1-score: 99.84\%, \newline AUC: 0.9990. & - \\ \hline
\cite{Transformners_5} & Availability & DoS, DDoS attacks & Network traffic data & End-user devices, communication networks, control systems, and data centers/cloud infrastructure & CICIDS2017 \newline CIC-DDoS2019 & CICIDS2017:\newline Accuracy: 99.35\%, \newline Precision: 0.9898, \newline Recall: 0.9883, \newline F1-Score: 99.17\%, \newline CIC-DDoS2019:\newline Accuracy: 98.58\%, \newline Precision: 0.9882, \newline Recall: 0.9866, \newline F1-Score: 98.48\%. & Performance drop on detecting SQL Injection. Challenges in choosing a threshold for high detection accuracy or low false alarm rate for specific traffic types. Less ideal performance on class types "Web-DDoS" and "SSDP" due to similar characteristics between them. \\ \hline
\cite{Transformners_2} & Integrity & Various network attack types & IoT sensors time-series data & End-user devices, communication networks, control systems, and data centers/cloud infrastructure & SWaT, \newline WADI, \newline SMAP, \newline MSL & SWaT: \newline F1-score: 91\%, \newline Recall: 0.881, \newline Precision: 0.7491. \newline WADI: \newline F1-score: 84\%, \newline Recall: 0.836, \newline Precision: 0.839 & Limitations in modeling long-term dependencies with recurrent mechanisms; less effective on datasets with weak feature dependencies like SMAP, MSL \\ \hline
\cite{Transformners_8} & Availability \newline Integrity \newline Confidentiality & Intrusion detection & Network traffic time-series data & End-user device, communication networks, control systems, data centers, and Cloud Infrastructure & KDD, \newline UNSW-NB15, \newline CICIDS-WED, \newline CICIDS-FRI & KDD: \newline Accuracy: 99.98\%, \newline F1-score: 99.97\%, \newline AUC: 0.9997 \newline UNSW-NB15: \newline Accuracy: 92.62\%, \newline F1-score: 95.44\%, \newline AUC = 0.8802 \newline CICIDS-WED: \newline Accuracy: 98.02\%, \newline F1-score: 98.15\%, \newline AUC: 0.9801 & Transformer-based self-supervised learning, masked context reconstruction, two-part loss function, robust to contamination, parameter sensitivity analysis \\ \hline
\cite{AE7} & Availability \newline Integrity \newline Confidentiality & Network Intrusions: zero-day attacks, DoS, DDoS, MITM, SQL Injection, Malware etc. & Network traffic: Protocol type, service type, flag status, port numbers, traffic stats: packet count, size, inter-arrival times, etc. & End-user device, communication networks and control systems & NSL-KDD, \newline IoTID20, \newline N-BaIoT & 
NSL-KDD: \newline MCC: 0.712, \newline F1-score: 89.5\%, \newline IoTID20: \newline  MCC: 0.595 \newline F1-score: 97.4\%
\newline N-BaIoT: MCC varies per device.
& Variable optimal model configurations and latent sizes. Threshold determination challenges. Limited model depth exploration; the need for further large-scale studies \\ \hline
\cite{RNN3} & Availability \newline Integrity \newline Confidentiality & Various security threats in SG environment & - & End user divide, control systems and networks & N-BaIoT & Accuracy: 99.73\%, \newline Precision: 0.973, \newline Recall: 0.9795, \newline F1-score: 97.56\% & - \\ \hline
\end{tabular}
\end{table*}

\subsection{NSL-KDD}
The NSL-KDD dataset is an improved version of the original KDDCUP99 dataset and is widely used as a benchmark dataset in network intrusion detection \cite{datasets_3}. It addresses limitations and evaluates the performance of IDS. This dataset contains essential records from the complete KDD dataset, ensuring sufficient records for comprehensive experiments.
The selection of records from each difficulty level is proportional to their percentage in the original dataset. Each record in the NSL-KDD dataset consists of 41 attributes representing different flow features and a label indicating the presence of an attack or normal behavior \cite{datasets_4}. The NSL-KDD dataset comprises various network traffic features extracted from a simulated environment, including TCP/IP connections with different attack types and normal behaviors. It provides separate training and test data files. The training dataset shares the same four attack types as the KDDCUP99 dataset\cite{datasets_5}. 
To overcome the limitations of the KDDCUP99 dataset, NSL-KDD eliminates redundant and unrealistic instances, and it offers a more balanced distribution of attack types, addressing the problem of high-class imbalance. This enhancement aims to produce an unbiased classifier and reduce false-positive results \cite{datasets_6}.
\subsection{UNSW-NB15}
The dataset was created and gathered by the Center for Cyber Security (ACCS) at the University of New South Wales (UNSW) in Canberra, Australia. It contains network traffic data representing recent normal activities and synthetic attacks \cite{datasets_2}. The data was collected by capturing raw network packets using the Tcpdump tool and processed with Argus, Bro-IDS tool, and 12 algorithms. This process extracted 49 features, including class labels such as source IP, destination IP, source port, destination port, transaction protocol, duration, and attack category. The dataset is divided into a training set (\texttt{UNSW\_NB15\_training-set.csv}) and a test set (\texttt{UNSW\_NB15\_testing-set.csv}). It consists of 43 attributes, including the class label, with ten categories, including one for normal activities and nine for different types of attacks. 
The UNSW-NB15 dataset offers a broader range of features, allowing for more effective feature selection than the KDD dataset, which serves as a comprehensive benchmark for evaluating IDS methods \cite{datasets_9}.

\subsection{CICIDS}
The CICIDS dataset by the Canadian Institutes for Cybersecurity is a valuable resource for evaluating intrusion detection systems, accurately reflecting real-world network traffic with 84 features per sample and covering 14 attack types. It captures distinct behavioral patterns of 25 users, facilitating multi-stage attack simulations. However, processing challenges arise due to fragmented data distribution. The CICIDS 2017 version addresses system complexity and dataset scarcity, creating diverse datasets for network-based anomaly detection. It includes user behavior profiles, abstract representations of network events, and seven attack scenarios, incorporating network traffic, system logs, and 80 extracted features. The 2018 version improves upon its predecessor by allowing more simulated clients and addressing class imbalance \cite{datasets_13}. The subsequent CSE-CIC-IDS-2019 dataset also enhances DDoS attack classification by including benign traffic and up-to-date attack scenarios\cite{datasets_14}.

\subsection{N-BaIoT}
The N-BaIoT dataset, available in the UCI {ML}Repository, comprises two well-known IoT botnets: BASHLITE (Gafgyt) and Mirai \cite{datasets_15}. It offers a significantly larger number of features compared to the NSL-KDD and CICIDS datasets. This dataset examines the characteristics of these botnets by collecting traffic data before and after infection, encompassing both normal and attack data from IoT botnet attacks. 
The dataset consists of 115 features. It represents the behavior of these botnets on a range of nine IoT devices, including thermostats, cameras, and doorbells. It provides traffic statistics for each node in a real IoT network, including packet sizes, data packet counts, and time intervals between packet arrivals.

\subsection{SGCC dataset}

The dataset examines the behavior of 42,372 customers from January 1, 2014, to October 31, 2016, using genuine electricity consumption data obtained from the State Grid Corporation of China (SGCC) \cite{datasets_16}. It encompasses these customers' daily kilowatt-hour (kWh) consumption records, comprising 38,757 regular users and 3,615 individuals flagged as electricity thieves. The dataset also includes ground truth information, revealing a severe issue in China where 9\% of customers commit electricity theft. The dataset has missing or inaccurate values that require preprocessing techniques due to various factors such as malfunctioning smart meters, imprecise measurement data, system maintenance, and data storage challenges. These missing values can misguide the classifier and lead to the misidentification of fraudulent consumers \cite{Others_LSTM_CNN3, Others_LSTM_CNN5}. It is worth mentioning that this dataset is imbalanced, posing a significant challenge in ETD, as it can lead to a biased classifier and hinder overall performance.

\subsection{Irish CER dataset}
The Irish CER Smart Metering Project dataset consists of load profiles from over 5000 residential users and small to medium-sized enterprises in Ireland. This dataset offers detailed information at 30-minute intervals collected over 500 days from July 2009 to December 2010. In addition to the smart meter data, a questionnaire was administered to participants, capturing data on various aspects such as occupant socio-demographic status, consumption behavior, household properties, and home appliances. The dataset includes 4,232 participants who actively took part in the research project. The questionnaire aimed to thoroughly comprehend households' attributes and actions concerning their electricity usage \cite{datasets_17}.

\section{Discussion}
\label{sec: Discussion}
{
DL methods have made significant progress in detecting cyber threats in SG, including sophisticated malware and DoS/DDoS attacks. However, their practical application faces challenges, particularly scalability and real-time data processing within the vast, dynamic SG networks. The models are particularly susceptible to rapid detection of threats, but their extensive computational requirements hinder their deployment in real-time environments. This indicates variability in the efficacy of the AE and One-Class Classification methodology, which is contingent upon the specific model selected—such as PAE, VAE, CNNAE(1D), CNNAE(2D)—and the volume of training data incorporated \cite{AE4}. As SG evolves, the demand for models to adapt to new connections and data without losing accuracy or requiring more resources intensifies. While techniques such as DDPG and Multiagent DRL show promise in threat detection and response, their performance in more extensive and diverse grid contexts is yet to be extensively validated \cite{DRL5, DRL8, DRL9}.}

{Models like LSTM, KFRNN, and CNN have effectively detected specific types of attacks within SG, including FDI attacks and energy theft. Nevertheless, since these models can adapt to various SG configurations and levels of complexity, their performance may vary. For example, the KFRNN architecture  \cite{RNN1} integrates unique features useful for proactive security measures by combining the temporal pattern recognition capabilities of RNN with the predictive capability of Kalman Filters. Additionally, evaluating advanced cybersecurity models like DCT-GAN, TGAN-AD, and RTIDS in SG cybersecurity is critical due to concerns about their effectiveness in controlled scenarios and lack of real-world operational testing. The complexity of specific models, such as GNNs for identifying FDI attacks, can present difficulties when deployed in real-time, potentially affecting their scalability and adaptability.}

{Models like RUIDS are efficient in SG cybersecurity due to their self-supervised learning capabilities, enabling reliable threat detection. However, determining how resilient these strategies are to varying attack vectors remains a challenge, particularly in the face of constantly changing and adapting to sophisticated cyber threats. Continuous improvement is crucial to overcome challenges like overfitting and assumptions in training data. Although AE show promise in feature selection, their reliability across different scenarios and attack vectors needs to be proven. They may face challenges in generalizing across various attack types, which could limit their scalability in diverse grid environments. Effective DL methods for SG must deliver consistent performance, adaptability to emerging threats, and evolve alongside the grid's advancements. Table \ref{tab:DL_details} provides a comprehensive analysis of cyber defense strategies in SG, emphasizing communication and power delivery networks. It categorizes cyberattacks, identifies specific aspects of SG infrastructure they target for security enhancement, and outlines the levels of SG operations susceptible to these attacks. Additionally, the table lists data features utilized in DL models for cyber defense and identifies security objectives at risk. It provides a thorough overview of the current status of SG cyber defense and addresses the drawbacks of these approaches.
}

{DeepMTD and Morphence enhance security in DL models against adversarial examples. However, the complexity involved in deploying several models after initial setup might pose challenges for practical applications. While Morphence's strategy of renewing its model pool proves effective against various attacks, relying on a predetermined query limit could lead to potential operational challenges. While "wAdvMTD" employs heterogeneous models to counter adversarial white-box examples, its accuracy may be compromised in complex tasks.
 A strategy uses physics-based MTD to counter FDI attacks in power systems. However, in high-noise environments, its probabilistic detection may be difficult. Another approach combines data-driven techniques with physics-based methods, but its implementation complexity could affect its effectiveness in real-time situations.}

{Additionally, models that utilize datasets such as Meraz'18 and Drebin lack specificity for the SG domain, affecting their relevance and applicability. While they detect known threats, their ability to generalize and recognize new cyber threats remains crucial for their applicability. Models showing high accuracy on generic datasets such as KDDCup99 and CICDDoS2019 do not guarantee effectiveness in SG environments. Tailored datasets such as SGCC and Irish CER are critical for training models to tackle the specific security challenges of SG. Consequently, datasets with current and varied attack vectors, like UNSW-NB15 and the more recent iterations of CICIDS, align with the necessities of modern IDS. The datasets create more accurate and realistic environments, enabling models to demonstrate their effectiveness in real-world scenarios.}

\section{Challenges and Future Work}
\label{sec: Challenges}
Several challenges emerge to enhance the security and robustness of {SG} through integrating MTD and DL techniques. These challenges and potential future directions are discussed in this section to provide a roadmap for researchers and practitioners. Specifically, using DL-based approaches, integrating MTD and DL techniques in {SG}, and applying various benchmark datasets present challenges and opportunities for future work.

\subsection{Challenges}
\paragraph{\textbf{Model Complexity}} The complexity and computational intensity of DL models can hinder real-time performance, particularly in {SG} where rapid response to threats is crucial. The challenge lies in developing lightweight yet effective models operating efficiently in real-time environments.
\paragraph{\textbf{Adversarial Attacks}} DL models are susceptible to adversarial attacks, which can significantly degrade their performance. The challenge is to enhance the robustness of these models against such attacks, which requires a deep understanding of adversarial strategies and the development of effective countermeasures.
\paragraph{\textbf{Data Quality}} The effectiveness of DL models relies heavily on the quality of the input data. Datasets with missing or inaccurate values, imbalanced classes, or outdated attack patterns can lead to biased or ineffective models. The challenge is to develop effective data preprocessing, cleaning, and augmentation techniques to improve the quality of input data.
\paragraph{\textbf{Dataset Specificity}} Most existing datasets are not specifically tailored for SG networks with unique characteristics and requirements. The challenge is to develop and maintain datasets that accurately reflect the dynamics of SG networks, including their normal operation and various types of attacks.

\subsection{Future Work}
\paragraph{\textbf{Model Optimization}} Future research could focus on optimizing DL models to reduce their complexity and computational requirements, making them more suitable for real-time SG applications. This could involve the development of new algorithms, the refinement of existing ones, or the application of techniques such as model pruning or quantization.
\paragraph{\textbf{Robustness Against Adversarial Attacks}} Developing new strategies and techniques to enhance the robustness of DL models against adversarial attacks is a promising area for future work. This could involve using adversarial training, defensive distillation, or other techniques.
\paragraph{\textbf{Data Preprocessing and Augmentation}} Techniques for effective data preprocessing and augmentation can help improve the quality of input data, leading to more accurate and reliable models. Future work could focus on developing new methods for data cleaning, imputation, and augmentation, as well as the application of techniques such as synthetic minority over-sampling (SMOTE) for addressing class imbalance.
\paragraph{\textbf{Dataset Development}} There is a need for datasets that are specifically tailored for SG networks. Future work could focus on developing such datasets, which would involve collecting network traffic data from real-world SG environments, simulating various types of attacks, and continuously updating the datasets to reflect evolving attack patterns.
\paragraph{\textbf{Real-world Testing}} More extensive testing of these techniques in real-world SG environments can help validate their effectiveness and identify areas for improvement. Future work could involve deploying these techniques in pilot SG networks, collecting performance data, and refining the techniques based on the results of these tests.

\section{Conclusion}
\label{sec: Conclusion}
In this paper, we explored how {SG} and the IIoT have increasingly relied on cutting-edge technologies, becoming enticing targets for intricate cyber-attacks. A comprehensive examination of the current state of proactive cyber defense strategies utilizing DL in {SG} and IIoT was conducted, as such a broad exploration was previously scarce in the literature.
We began with an overview of related works and our unique contributions, delving into an examination of SG infrastructure. Various cyber defense techniques were classified into reactive and proactive categories, with a marked emphasis on DL-enabled proactive defenses. We also briefly discussed the intricacies of IDS. {Subsequently, we highlight potential vulnerabilities and corresponding security objectives, underscoring the importance of continuous monitoring and proactive risk mitigation against evolving cyber threats.}
A thorough taxonomy of DL approaches was provided, underlining their functions and significance in the proactive protection of {SG} and IIoT. We analyzed the most substantial DL-based methods currently in use and explored the interaction between Moving Target Defense, a proactive defense strategy, and DL methodologies. The inclusion of an overview of benchmark datasets substantiated the discourse.
{The advancements and limitations of sophisticated approaches in cyber threat detection are then discussed, addressing concerns such as scalability and real-time reaction in dynamic network settings.}
The paper concluded by addressing the challenges of implementing DL-based security systems within {SG} and IIoT and providing a forward-looking perspective on future developments in this crucial field. The insights gleaned underscored the potential of DL in transforming the security landscape of {SG} and IIoT, highlighting the necessity for ongoing innovation and multidisciplinary collaboration.

\bibliographystyle{IEEEtran}
\bibliography{Reference}
\begin{IEEEbiography} [{\includegraphics[width=1in,height=1.25in,clip,keepaspectratio]{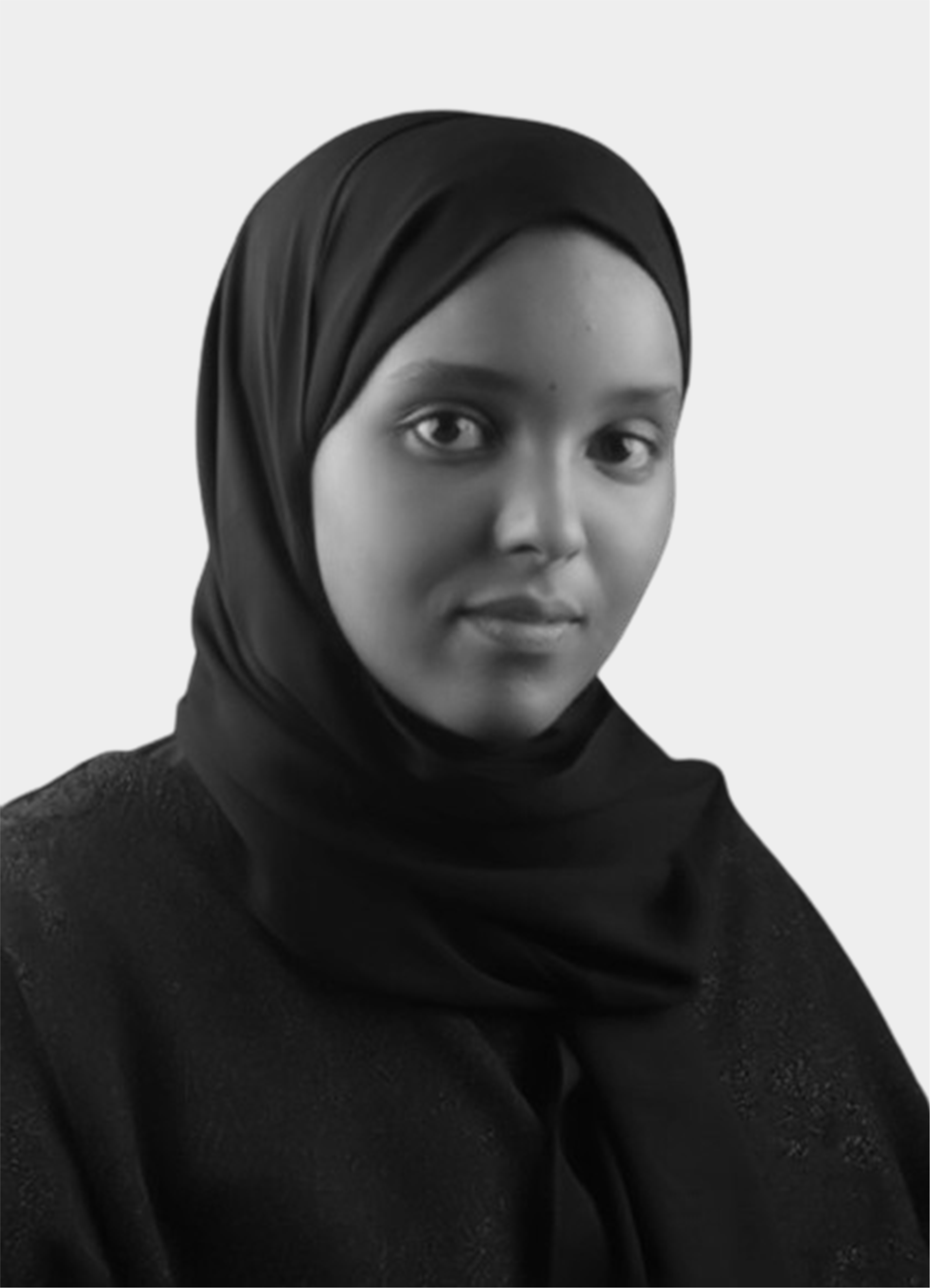}}]  { \rmfamily  Nima Abdi}
\rmfamily \mdseries received her B.Sc. in Electrical Engineering from Qatar University in 2020 and is currently pursuing an M.Sc. in Data Science and Engineering at Hamad Bin Khalifa University (HBKU). Her research focus is on the application of Artificial Intelligence on smart grid security, specifically the physical layer.
\end{IEEEbiography}

\begin{IEEEbiography} [{\includegraphics[width=1in,height=1.25in,clip,keepaspectratio]{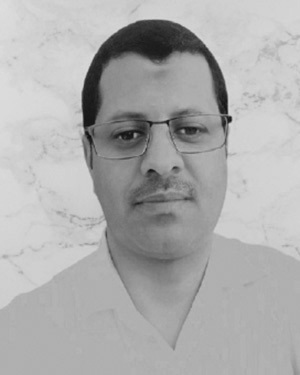}}] { \rmfamily  Abdullatif Albaseer (Member, IEEE)} \rmfamily \mdseries received an M.Sc. degree in computer networks from King Fahd University of Petroleum and Minerals, Dhahran, Saudi Arabia, in 2017 and a Ph.D. degree in computer science and engineering from Hamad Bin Khalifa University, Doha, Qatar, in 2022. He is a Postdoctoral Research Fellow with the Smart Cities and IoT Lab at Hamad Bin Khalifa University. He has authored and co-authored more than twenty conference and journal papers in IEEE ICC, IEEE Globecom, IEEE CCNC, IEEE WCNC and IEEE Transactions. He also has six US patents in the area of the wireless network edge. His current research interests include AI for Networking, AI for Cybersecurity, Distributed AI, and Edge Intelligence.
\end{IEEEbiography}
\begin{IEEEbiography} [{\includegraphics[width=1in,height=1.25in,clip,keepaspectratio]{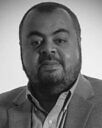}}]
{ \rmfamily  Mohamed Abdallah (Senior Member, IEEE)} \rmfamily \mdseries
received his B.Sc. degree from Cairo University, Giza, Egypt, in 1996, and his M.Sc. and Ph.D. degrees from the University of Maryland at College Park, College Park, MD, USA, in 2001 and 2006, respectively.,From 2006 to 2016, he held academic and research positions with Cairo University and Texas A \& M University in Qatar, Doha, Qatar. He is currently a Founding Faculty Member with the rank of Associate Professor with the College of Science and Engineering, Hamad Bin Khalifa University, Doha. He has published more than 150 journals and conferences and four book chapters and co-invented four patents. His current research interests include wireless networks, wireless security, smart grids, optical wireless communication, and blockchain applications for emerging networks. He is a recipient of the Research Fellow Excellence Award at Texas A\& M University in Qatar in 2016, the Best Paper Award in multiple IEEE conferences, including IEEE BlackSeaCom 2019 and the IEEE First Workshop on Smart Grid and Renewable Energy in 2015, and the Nortel Networks Industrial Fellowship for five consecutive years, 1999–2003. His professional activities include an Associate Editor of the IEEE Transactions on Communications and the IEEE Open Access Journal of Communications, the Track Co-Chair of the IEEE VTC Fall 2019 Conference, the Technical Program Chair of the 10th International Conference on Cognitive Radio-Oriented Wireless Networks, and a technical program committee member of several major IEEE conferences.
\end{IEEEbiography}
\end{document}